\documentclass[aps,prb,twocolumn,showpacs,floatfix]{revtex4}
\usepackage{graphicx}
\begin{document}
\newcommand{\App}{A^{\prime\prime}}
\title{Interactions and dynamics in Li + Li$_2$ ultracold collisions}
\author{Marko T. Cvita\v{s}\footnote{Present address: University
Chemical Laboratory, Lensfield Road, Cambridge, CB2 1EW, UK}}
\author{Pavel Sold\'{a}n\footnote{Present address: Doppler Institute,
Department of Physics, Faculty of Nuclear Sciences and Physical
Engineering, Czech Technical University, B\v{r}ehov\'{a} 7, 115 19
Praha 1, Czech Republic}}
\author{Jeremy M. Hutson}
\affiliation{Department of Chemistry, University of Durham, South
Road, Durham, DH1 3LE, UK}
\author{Pascal Honvault\footnote{Present address: Institut UTINAM,
UMR CNRS 6213, Universit\'e de Franche-Comt\'e, 25030 Besan\c{c}on
Cedex, France}}
\author{Jean-Michel Launay}
\affiliation{UMR 6627 du CNRS, Laboratoire de Physique des Atomes,
Lasers, Mol\'{e}cules et Surfaces, Universit\'{e} de Rennes, 35042
Rennes Cedex, France}
\begin{abstract}
A potential energy surface for the lowest quartet electronic state
($^4A'$) of lithium trimer is developed and used to study
spin-polarized Li + Li$_2$ collisions at ultralow kinetic
energies. The potential energy surface allows barrierless atom
exchange reactions. Elastic and inelastic cross sections are
calculated for collisions involving a variety of rovibrational
states of Li$_2$. Inelastic collisions are responsible for trap
loss in molecule production experiments. Isotope effects and the
sensitivity of the results to details of the potential energy
surface are investigated. It is found that for vibrationally
excited states the cross sections are only quite weakly dependent
on details of the potential energy surface.
\end{abstract}
\maketitle

\section{Introduction}

There is great interest in the properties of molecules formed in
laser-cooled atomic gases by processes such as photoassociation
and magnetic association. \cite{Hutson:IRPC:2006, Jones:RMP:2006,
Koehler:RMP:2006} Diatomic molecules have been formed from bosonic
and/or fermionic species of all the alkali metals from Li to Cs.
Dimers of bosonic species have usually been found to decay quickly
because of inelastic collisions with other atoms,
\cite{Donley:2002, Herbig:2003, Xu:2003, Durr:mol87Rb:2004} but
for fermionic species ($^6$Li, $^{40}$K) it has been possible to
stabilise the dimers by tuning to large positive values of the
atom-atom scattering length. \cite{Strecker:2003, Cubizolles:2003,
Jochim:Li2pure:2003} This approach was used in late 2003 to create
long-lived Bose-Einstein condensates of fermion dimers,
\cite{Jochim:Li2BEC:2003, Zwierlein:2003, Greiner:2003} and since
that time there has been a large amount of work on their
properties, and particularly on the transition between
Bose-Einstein condensation (BEC) and the Bardeen-Cooper-Schrieffer
(BCS) regime characterised by long-range Cooper pairing and
superfluidity.

Photoassociation and magnetic association (tuning through
zero-energy Feshbach resonances) both produce molecules that are
initially in very highly excited vibrational states. However,
there is great interest in bringing these molecules down to low
vibrational states and ultimately to the absolute ground state.
For example, Staanum {\em et al.} \cite{Staanum:2006} have
produced Cs$_2$ in vibrational levels $v=4$ to 6 of the
$a^3\Sigma^+$ state and studied collision processes, while Sage
{\em et al.} \cite{Sage:2005} have produced small numbers of
ultracold polar RbCs molecules in their absolute ground state by a
4-photon process.

Ultracold molecules are initially formed in the presence of
ultracold atoms and can collide with them. \cite{Hutson:IRPC:2007}
For molecules in vibrationally excited states, there is the
possibility of vibrationally inelastic collisions,
\begin{equation} \hbox{M}_2(v) + \hbox{M} \longrightarrow
\hbox{M}_2(v^\prime<v) + \hbox{M}, \end{equation} where $v$ is the
vibrational quantum number. Since the trap depth is usually much
less than 1 K, such collisions always release enough kinetic
energy to eject both collision partners from the trap. If the
molecular density is high, there is also the possibility of
inelastic molecule-molecule collisions,
\begin{equation} \hbox{M}_2(v) + \hbox{M}_2(v) \longrightarrow
\hbox{M}_2(v^\prime<v) + \hbox{M}_2 (v^{\prime\prime}\le v).
\end{equation}
Molecules are not {\it destroyed} in inelastic collisions, but
they are lost from the trap and are no longer ultracold.

The rates of inelastic atom-molecule collisions involving alkali
metal dimers have been studied both experimentally
\cite{Wynar:2000, Mukaiyama:2004, Staanum:2006, Zahzam:2006} and
theoretically. \cite{Soldan:2002, Quemener:2004, Petrov:2004,
Petrov:suppress:2005, Quemener:2005, Cvitas:bosefermi:2005,
Cvitas:hetero:2005} Wynar {\em et al.} \cite{Wynar:2000} formed
$^{87}$Rb$_2$ molecules in the second-to-last vibrational level of
the ground excited state by stimulated Raman adiabatic passage
(STIRAP). They estimated an upper bound of $k_{\rm
loss}=8\times10^{-11}$ cm$^3$ s$^{-1}$ due to inelastic
atom-molecule collisions. Mukaiyama {\em et al.}
\cite{Mukaiyama:2004} measured the trap loss rate for
$^{23}$Na$_2$ molecules formed by Feshbach resonance tuning and
obtained an atom-molecule rate coefficient $k_{\rm loss} =
5.1\times10^{-11}$ cm$^3$ s$^{-1}$ for molecules in the highest
vibrational state. Staanum {\em et al.} \cite{Staanum:2006}
investigated inelastic collisions of rovibrationally excited
Cs$_2$ ($^3\Sigma_u^+$) in collisions with Cs atoms in two
different ranges of the vibrational quantum number $v$ by
monitoring trap loss of Cs$_2$. They obtained atom-molecule rate
coefficients close to $1.0\times10^{-10}$ cm$^3$ s$^{-1}$ for both
$v=4$ to 6 and $v=32$ to 47. Zahzam {\em et al.}
\cite{Zahzam:2006} carried out similar work for different
rovibrational states of $^3\Sigma_u^+$, and also considered
molecules in the $^1\Sigma_g^+$ state and molecule-molecule
collisions. They obtained rate coefficients of $2.6\times10^{-11}$
cm$^3$ s$^{-1}$ and $1.0\times10^{-11}$ cm$^3$ s$^{-1}$ in the
atom-atom and atom-molecule cases respectively, both with quite
large error bounds.

Quantum dynamics calculations on alkali metal atom--diatom
collisions were first carried out by Sold\'{an} {\em et al.},
\cite{Soldan:2002} using the potential energy surface of Higgins
{\em et al.} \cite{Higgins:2000} and subsequently extended to K +
K$_2$ on a new potential surface. \cite{Quemener:2005} In parallel
work, Petrov {\em et al.} \cite{Petrov:2004, Petrov:suppress:2005}
analysed the stability of fermion dimers in terms of the
long-range form of the wavefunction. In the case where the
atom-atom scattering length $a$ is much larger than the range of
the atom-atom potential $r_e$, they showed that both atom-molecule
and molecule-molecule inelastic collision rates are suppressed by
Fermi statistics. However, their derivation applies only to
molecules that are in long-range states, with a wavefunction that
depends on the atom-atom scattering length $a$, with $\chi(r) \sim
\exp(-r/a)$. In the present paper, we show computationally that
there is {\em no} systematic suppression of the atom-molecule
inelastic rate for $^6$Li dimers in low-lying vibrational levels,
even when $a$ is large and positive. We also consider
mixed-isotope Li + Li$_2$ collisions. This has been reported
briefly in previous work, \cite{Cvitas:bosefermi:2005,
Cvitas:hetero:2005} but the present paper gives full details of
the calculations and further details of the results.

The structure of the paper is as follows. Section II describes
calculations of the potential energy surface for quartet Li$_3$,
including details of both the electronic structure methods
employed and the procedures used to interpolate between and
extrapolate beyond the {\em ab initio} points. The surface allows
barrierless atom exchange reactions. Section III describes atom +
diatom reactive scattering calculations both for homonuclear
systems [bosonic, $^7$Li + $^7$Li$_2$ and fermionic, $^6$Li +
$^6$Li$_2$] and mixed-isotope systems. For collisions of $^7$Li
with either $^6$Li$_2$ or $^6$Li$^7$Li, exoergic atom exchange
reactions are possible even for ground-state molecules because of
the differences in zero-point energies. Section III also explores
the sensitivity of the cross sections to the potential energy
surface and shows that for vibrationally excited states the
dependence is relatively weak (only a factor of 2 for molecules
initially in $v=3$). Section IV presents conclusions and prospects
for future work.

\section{Quartet potential energy surfaces for \protect{Li$_3$}}

In previous work, we investigated nonadditive forces in
spin-polarized (quartet) alkali metal trimers \cite{Soldan:2003}
at linear and equilateral geometries. For Li$_3$, we found very
large nonadditive forces at equilateral geometries that reduced
the interatomic distance by more than 1 \AA\ and increased the
well depth by a factor of 4 with respect to the sum of Li--Li pair
potentials. We subsequently gave a brief description of the
complete potential energy surface \cite{Cvitas:bosefermi:2005} and
identified a seam of conical intersections at collinear geometries
between $^4\Sigma$ and $^4\Pi$ states. The conical intersection
has been investigated further by Brue {\em et al.}
\cite{Brue:2005} The seam results in a cusp at energies close to
the atom-diatom threshold and influences Li + Li$_2$ collision
dynamics at ultralow energies. \cite{Cvitas:bosefermi:2005,
Cvitas:hetero:2005} In the present paper, we give a full
description of the {\em ab initio} calculations and fitting
procedure used to obtain a global representation of the surface
and extend the collision calculations.

\subsection{Electronic states overview}

In the present work, we are interested mostly in the lowest
quartet electronic state of Li$_3$, which is a $^4A'$ state in
$C_s$ symmetry. This state correlates with three ground-state
($^2S$) atoms at large interatomic separations. As mentioned
above, the quartet ground state has crossings with other
electronic states that correlate with the $S+S+P$ separated-atom
limit. These crossings (and avoided crossings) occur at energies
that are potentially relevant to scattering processes involving
three ground-state Li atoms. To gain a broader, qualitative
picture on how the different potential energy surfaces intertwine,
we have carried out state-averaged multiconfiguration
self-consistent field (MCSCF) calculations of the potential energy
surfaces of quartet states for certain high-symmetry arrangements.
These calculations used the complete active space (CASSCF) method
of Werner and Knowles, \cite{Werner:1980, Werner:1981,
Werner:1985, Knowles:1985, Werner:1987} as implemented in the
MOLPRO package, \cite{MOLPRO} with an aug-cc-pVTZ basis set
generated from the cc-pVTZ set \cite{EMSL} as described below. The
active space included all 24 molecular orbitals constructed
principally from 2s, 2p, 3s and 3p atomic orbitals, with the three
1s orbitals doubly occupied in all configurations; this active
space is designated [3,24] below.

\begin{figure} [tbp]
\begin{center}
\includegraphics[width=60mm,angle=-90]{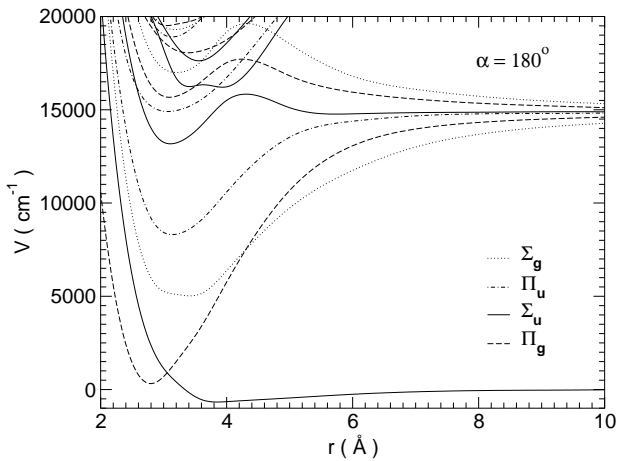}
\caption{CASSCF quartet potentials of Li$_3$ at $D_{\infty h}$
geometries for states that correlate with the atomic $S + S + S$
and $S + S + P$ asymptotic limits. The interatomic distances are
$r$, $r$, and $2r$.} \label{fig1}
\end{center}
\end{figure}

\begin{figure} [tbp]
\begin{center}
\includegraphics[width=60mm,angle=-90]{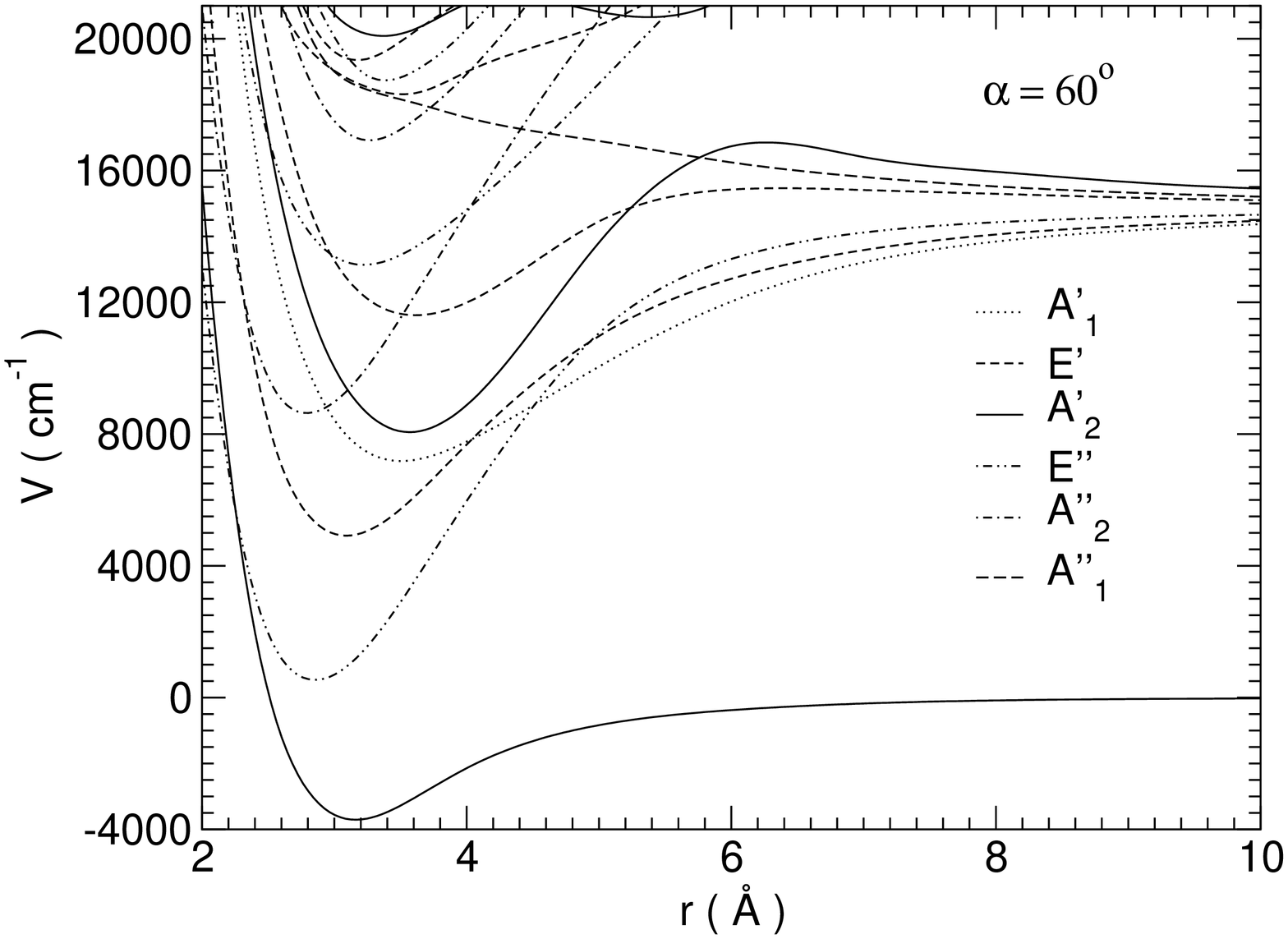}
\caption{CASSCF quartet potentials of Li$_3$ at $D_{3 h}$
geometries for states that correlate with the atomic $S + S + S$
and $S + S + P$ asymptotic limits. The interatomic distances are
$r$.} \label{fig2}
\end{center}
\end{figure}

Figure \ref{fig1} shows the potential curves obtained for quartet
states at $D_{\infty h}$ geometries. The ground state is of
$\Sigma_u$ symmetry; at a Li-Li bond length $r$ near 3 \AA\ it
intersects a $\Pi_g$ state, which in turn intersects a $\Sigma_g$
state at higher energies. If we change the interbond angle
$\alpha$, keeping the two adjacent bond lengths unchanged (moving
from $D_{\infty h}$ to $C_{2v}$ arrangements), the ground state
becomes of $B_2$ symmetry, while the $\Pi_g$ state splits into
states of $A_2$ and $B_2$ symmetries. The two $B_2$ states
avoided-cross around $\alpha=70^\circ$. At $D_{3h}$ arrangements
($\alpha=60^\circ$), shown in Figure \ref{fig2}, the ground state
is of $A'_2$ symmetry, while the second $B_2$ state forms a doubly
degenerate $E'$ state with the $A_1$ state that originated from
$\Sigma_g$ at $D_{\infty h}$. The $A_2$ state in $C_{2v}$ symmetry
($\Pi_g=A_2+B_2$) forms an $E''$ state with the $B_1$ state that
was a part of $\Pi_u=A_1+B_1$ at $D_{\infty h}$.

\begin{figure} [tbp]
\begin{center}
\includegraphics[width=60mm,angle=-90]{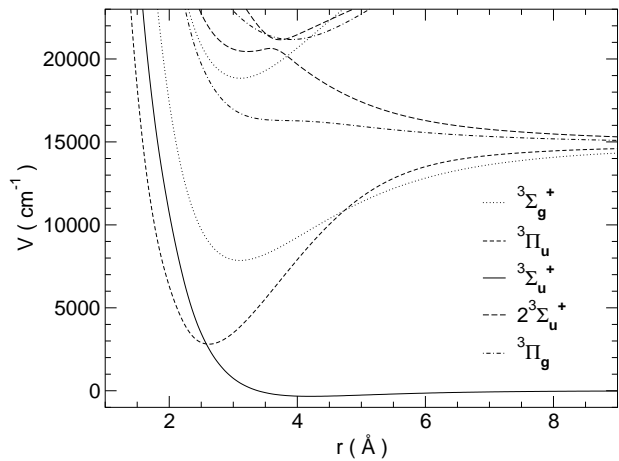}
\caption{Triplet potential energy curves of Li$_2$ from the atomic
$S + S$ and $S + P$ dissociation limits.} \label{fig3}
\end{center}
\end{figure}

For comparison, Figure \ref{fig3} shows the triplet states of
Li$_2$ calculated using [2,8]-CASSCF, followed by an internally
contracted multireference configuration interaction (MRCI)
\cite{Werner:1982, Werner:1987, Knowles:1988, Werner:1988}
including single and double excitations from the CASSCF
wavefunction.

\begin{figure} [tbp]
\begin{center}
\includegraphics[width=70mm,angle=-90]{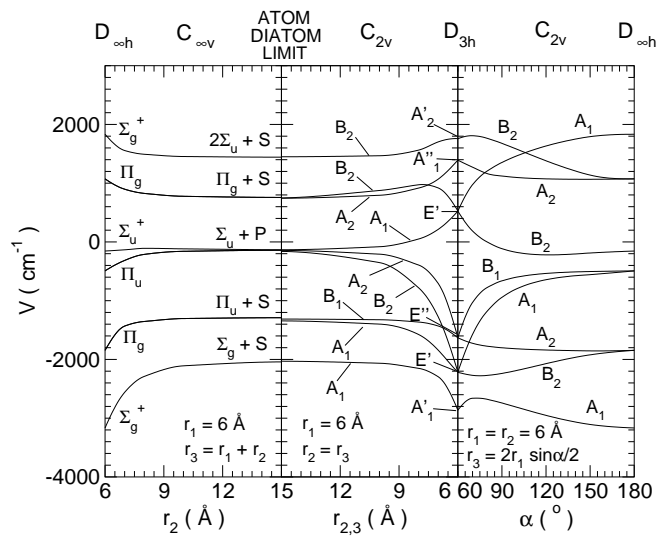}
\caption{Correlation diagram of quartet potentials of Li$_3$ that
correlate with the atomic $S + S + P$ dissociation limit. Note the
the energy is shown relative to this limit and the state
correlating with $S+S+S$ is not shown. The first panel connects
the $D_{\infty h}$ states with the atom-diatom limit, with one
interatomic distance fixed at 6 \AA \ and $C_{\infty v}$ symmetry
preserved. The second panel connects the atom-diatom limit with
$D_{3h}$ states, with one interatomic distance fixed at 6 \AA \
and $C_{2v}$ symmetry preserved. The third panel connects $D_{3h}$
and $D_{\infty h}$ terms, with two interatomic distances fixed at
6 \AA \ and the angle between them, $\alpha$, varied.}
\label{fig:correl}
\end{center}
\end{figure}

Figure \ref{fig:correl} shows a correlation diagram that connects
the $D_{\infty h}$, $D_{3h}$ and atom-diatom ($C_{2v}$) limits for
states of quartet Li$_3$ that correlate with the $S+S+P$ atomic
limit for fixed bond length of 6 \AA. Note that the molecular
state arising from three $S$-state atoms is not shown.

In the present work we focus on calculating the lowest quartet
state, $^4A'$, to high precision. Colavecchia {\em et al.}
\cite{Colavecchia:Li3:2003} and Brue {\em et al.} \cite{Brue:2005}
have carried out full configuration interaction (FCI) calculations
with just one electron on each Li atom correlated. Our
calculations differ from theirs in correlating all electrons and
in using considerably larger basis sets, but use approximate
correlation treatments.

\subsection{Basis set convergence}

A variety of basis sets are available for lithium. Feller
\cite{EMSL} constructed correlation-consistent polarized valence
basis sets (cc-pVXZ for X$=$D, T, Q, and 5) \cite{Dunning:1989}
and Iron {\em et al.} \cite{Iron:2003} devised
correlation-consistent polarized core-valence basis sets
(cc-pCVXZ) by adding ``tight" functions to Dunning-type cc-pVXZ
bases in order to give a better account of the core-core and
core-valence correlation. There are no standard augmented basis
sets available for Li, \cite{EMSL} so we generated augmented
variants of several basis sets using the even-tempered scheme
implemented in MOLPRO, which adds an additional diffuse function
of each angular symmetry (s, p, d, f, ...), with an even-tempered
exponent based on the ratio of the two smallest exponents in the
original set (or a ratio of 2.5 if only one function of a type is
present in the original basis).

The cc-pVXZ and cc-pCVXZ basis sets are usually used with the s
and p functions contracted. However, in previous work on alkali
metal trimers, \cite{Soldan:2003} we found it advantageous to use
the basis sets in uncontracted form. In the present work we
consider the cc-pV5Z basis set in three different forms: fully
contracted (sp-contracted), fully uncontracted, and with just the
p space uncontracted (s-contracted). However, we found it too
expensive to uncontract the cc-pCVXZ and augmented basis sets and
these were used in fully contracted form.

We chose to use RCCSD(T) calculations (restricted open-shell
coupled cluster with single, double and noniterative triple
excitations). All calculations were carried out using MOLPRO.
\cite{MOLPRO}

With the computer resources available to us, we found that
RCCSD(T) calculations of triatomic energies with the largest basis
sets available, cc-pCV5Z and aug-cc-pCV5Z, were feasible only for
relatively small numbers of points, not for a grid of several
hundred points. These basis sets were therefore used only for
atomic and diatomic calculations, except that the cc-pCV5Z basis
set was used to provide benchmark triatomic calculations at a
restricted number of points.

\begin{table*}[tbp]
\begin{center}
\begin{tabular}{lccccccccc} \cline{1-10}
      &  Atomic         & & Diatomic    &&&&& Triatomic \\
basis set & $\alpha(a_0^3)$ & $S$--$P$ energy & $r_e$ (\AA) &
$D_e$ & $a$ ($a_0$)
      & $E_{10}$    & BSSE & $r_e$ (\AA) & $D_e$ \\
      & & (cm$^{-1}$) && (cm$^{-1}$) && (cm$^{-1}$) & (cm$^{-1}$) & & (cm$^{-1}$) \\
\cline{1-10}

sp-contracted cc-pV5Z & 165.518 & 14834.01 & 4.169 & 334.042 &          &         & 41.45 & 3.102 & 4021.5 \\
s-contracted cc-pV5Z  & 164.396 & 14903.62 & 4.177 & 328.922 & $12.54$  & $0.223$ & 2.553 & 3.102 & 3969.8 \\
uncontracted cc-pV5Z  & 164.336 & 14906.69 & 4.177 & 328.732 &          &         & 0.8015 & 3.102 & 3967.4 \\
aug-cc-pCV5Z          & 164.189 &          & 4.176 & 330.548 & $-8.95$  & $0.346$ & 0.234 \\

cc-pCV5Z              & 164.152 & 14906.48 & 4.175 & 328.952 &          &         & 0.139 & 3.098 & 3980.6 \\

cc-pCVQZ              & 164.140 & 14911.28 & 4.179 & 325.499 &          &         & 0.501 \\


Best available theory& 164.111$^a$ \\

Experiment     & 164.0$\pm$3.4$^b$  & 14903.89$^c$ & 4.173 & $333.69$  & $-27.3\pm0.8$  & $0.416$ \\
\hline
\end{tabular}
\vspace{0.5cm} \caption{Convergence tests for Li basis sets. For
atoms: static polarizability $\alpha$ and $S$--$P$ excitation
energy. For triplet dimers: dissociation energy $D_e$, position of
the minimum $r_e$, BSSE evaluated at $r=4.2$~\AA\, scattering
length $a$ and energy of the highest vibrational level $E_{10}$
for the $^7$Li$_2$ molecule. For quartet trimers, dissociation
energy $D_e$ and position of the minimum $r_e$ for Li$_3$
($D_{3h}$). All calculations except ``best available theory" are
from the present work and used RCCSD(T) calculations. $^a$: ref.\
\onlinecite{Yan:1996}; $^b$: ref.\ \onlinecite{Molof:1974}; $^c$:
ref.\ \onlinecite{Moore:1971}.} \label{tab1}
\end{center}
\end{table*}

We began by testing convergence of two atomic properties, the
static polarizability and the $S$--$P$ excitation energy, which
are important for accurate calculations of long-range forces. The
results for a variety of basis are shown in Table \ref{tab1}. In
general terms all the basis sets perform acceptably for both
quantities, though the s-contracted cc-pV5Z basis set is
fortuitously accurate for the $S$--$P$ excitation energy.

We next considered the performance of different basis sets on
potential curves for the $^3\Sigma_u^+$ state of Li$_2$.
Electronic energies were calculated on a mesh of 64 interatomic
distances.
The results were corrected for basis-set superposition error
(BSSE) using the counterpoise correction. \cite{Boys:1970} The
points were then interpolated using the reciprocal power
reproducing kernel Hilbert space (RP-RKHS) interpolation method
\cite{Ho:1996,Ho:2000} with dispersion coefficients $C_6$, $C_8$
and $C_{10}$ fixed to accurate values from the calculations of Yan
{\em et al.} \cite{Yan:1997} At internuclear distances $r>16$ \AA\
the {\em ab initio} energies were replaced by those from the
three-term dispersion formula. Other interpolation parameters, in
the notation of Ref.~\onlinecite{Ho:2000}, were $m=2$, $n=3$,
$r_a=15$ \AA, $r_{65}=21.5$ \AA, $r_{66}=22.5$ \AA, $r_{67}=23.5$
\AA.


The resulting diatomic well depth $D_e$ and bond length $r_e$ are
included in Table \ref{tab1}, together with the binding energy of
the last bound state $E_{10}$ and scattering length $a$ calculated
for $^7$Li$_2$. The results are compared with properties obtained
from the RKR curves of Linton {\em et al.} \cite{Linton:1999} (the
values of $a$ and $E_{10}$ are from photoassociation spectroscopy
in combination with RKR results \cite{Abraham:1995}). It may be
seen that all the basis sets underestimate the RKR well depth by
between 2 and 5 cm$^{-1}$, except for the cc-pCVQZ basis set which
underestimates it by 8 cm$^{-1}$ and the sp-contracted cc-pV5Z
basis set which slightly overestimates it. On this basis we
excluded the cc-pCVQZ basis set from further consideration.

Another way to test the quality of a basis set is to consider the
magnitude of BSSE. BSSE arises when basis functions on two centers
overlap and compensate for inadequacies in the one-centre basis.
Table \ref{tab1} shows dimer BSSE values at $r=4.2$~\AA, which is
near the potential minimum. The cc-pCVXZ basis sets give quite
small BSSE values since they are by construction geared to
represent the core-core correlation well. However, as discussed
above, the cc-pCVQZ basis set is not accurate enough for our
purposes and the cc-pCV5Z basis set is too expensive. We therefore
focussed on the cc-pV5Z basis sets. As may be seen in Table
\ref{tab1}, the cc-pV5Z basis set in sp-contracted form gives a
very large BSSE, 41.45 cm$^{-1}$ at $r=4.2$ \AA, which we
considered unacceptably high. However, uncontracting the p space
reduces this to 2.553 cm$^{-1}$. Uncontracting the s space as well
reduces this by a further factor of 3, but at a very considerable
computational cost.

Finally, we carried out convergence tests on triatomic energies.
In this case there are no experimental results to compare with, so
we used the position and energy of the equilateral minimum,
calculated with the cc-pCV5Z basis set, as our benchmark. As may
be seen in Table \ref{tab1}, there is very little difference
between the s-contracted and uncontracted forms of the cc-pV5Z
basis set, which underestimate the cc-pCV5Z well depth by 11 and
13 cm$^{-1}$ respectively. However, the sp-contracted form
overestimates the well depth by 40 cm$^{-1}$. For triatomic
RCCSD(T) calculations the s-contracted basis set costs only 10\%
more than the sp-contracted set, which the fully uncontracted set
costs twice as much. On this basis we decided to proceed with the
s-contracted cc-pV5Z basis set to calculate the full potential
surface.

One further possibility we considered was the use of frozen-core
calculations, which are considerably cheaper than calculations
that correlate all electrons. However, we found that such
calculations systematically overestimate dimer well depths by 2 to
3\% and equilibrium distances by about 2\% compared to cc-pCVXZ
calculations at each level (T, Q, Z). For example, frozen-core
calculations with the sp-contracted cc-pV5Z basis set give
$D_e=338.471$ cm$^{-1}$ and $r_e=4.199$ \AA\ for Li$_2$.
Frozen-core calculations on the trimer with this basis set {\em
underestimate} the trimer well depth, giving $D_e=3937.5$
cm$^{-1}$ and $r_e=3.137$ \AA. We therefore did not pursue the
frozen-core approximation.


As described above, the Li$_3$ potential shows a seam of conical
intersections at linear geometries. The RCCSD(T) approach is
inherently a single-reference method, so its appropriateness in
the vicinity of a conical intersection is open to question. The
usual approach is to consider the T1 diagnostic, which can be used
to identify where multireference effects become large and may
compromise the results. Lee and Taylor \cite{Lee:1989} have given
a rule of thumb that single-reference methods become unreliable
when $\mbox{T1} > 0.02$. In the present calculations, the T1
diagnostic was relatively small at near-linear geometries but had
a maximum value of around 0.11 at equilateral geometries. We
therefore investigated the reliability of the RCCSD(T) approach by
recalculated the Li$_3$ potential energy surface with a smaller
basis set (aug-cc-pVTZ), using [3,12]-CASSCF followed by a
multireference configuration interaction (MRCI) calculation
including Davidson's correction, and compared it with the
corresponding RCCSD(T) energies. The T1 diagnostic was again large
in the well region for bond angles of $\sim 60^\circ$.
Nevertheless, the two surfaces agreed within 2\% in well depth
with no visible qualitative differences. We therefore believe that
the RCCSD(T) energies obtained with the cc-pV5Z basis set are
trustworthy.


\subsection{The $1^4A'$ surface}
\subsubsection{Representation of the surface}

If the potential is to be capable of representing all the
properties of experimental interest (including atom-atom
scattering lengths, dimer and trimer bound states, atom-diatom
collisions and 3-body recombination), then it is very important
that it should represent dissociation into all possible sets of
products (atom + diatom and 3 separated atoms) with the correct
long-range behaviour. In our previous work on K + K$_2$,
\cite{Quemener:2005} this was achieved by separating the potential
into pairwise-additive and non-additive parts and interpolating
them separately,
\begin{equation}
\label{Eqnonad} V({\bf r}) = \sum_{i<j}^3 V_{\rm dimer}(r_{ij}) +
V_{3}({\bf r}),
\end{equation}
where {\bf r} represents $(r_{12},r_{23},r_{31})$ and the
functional form for the nonadditive part $V_{3}({\bf r})$ is
chosen to ensure the correct behaviour in the atom-diatom limit.
\cite{Cvitas:longrange:2006} However, for Li + Li$_2$ the
non-additive term is so large that the potential minimum for the
trimer occurs at a distance that is high on the repulsive wall for
the dimer. Because of this, Eq.\ (\ref{Eqnonad}) would represent
the interaction potential in this region as a difference between
two very large quantities. This is unsatisfactory. Nevertheless,
at long range a decomposition according to Eq.\ (\ref{Eqnonad}) is
essential. Under these circumstances, it is best to fit the {\it
ab initio} points directly to obtain a short-range function
$V_{\rm SR}({\bf r})$ without imposing the correct long-range
behaviour, and to join this onto a potential with the proper
long-range behaviour $V_{\rm LR}({\bf r})$ using a switching
function $S({\bf r})$,
\begin{equation}
V({\bf r}) = S({\bf r}) V_{\rm SR}({\bf r}) + [1-S({\bf r})]
V_{\rm LR}({\bf r}).
\end{equation}
The switching function is 1 at short range but switches smoothly
to zero at long range as described below.

\subsubsection{Choice of grid}

For quantum dynamics calculations, it is very important to
represent the potential energy function smoothly and without
oscillations between \textit{ab initio} points. There are several
coordinate systems that can be used for triatomic systems,
including hyperspherical coordinates, Jacobi coordinates, valence
(bond length / bond angle) coordinates and pure bond-length
coordinates. These are by no means equivalent for interpolation
purposes. As will be seen below, the dynamical calculations are
carried out in hyperspherical coordinates. However, grids of
points in hyperspherical coordinates tend to include points in
which 2 atoms lie very close together, which hinders interpolation
because polynomials with very high localised maxima tend to have
oscillations in other regions. Jacobi coordinates suffer from the
same problem, and neither Jacobi coordinates nor valence
coordinates allow the full 3-body exchange symmetry to be
introduced in a natural way. Fortunately, there is no need to
represent the potential energy surface in the same coordinate
system as is used in the dynamical calculations. We therefore
chose to carry out electronic structure calculations on a grid of
points in pure bond-length coordinates $(r_{12},r_{23},r_{31})$,
which make exchange symmetry easy to handle but avoid points with
atoms too close together.

The grid was constructed using the internuclear distances $r=2.0$,
2.4, 2.8, 3.2, 3.6, 4.0, 4.4, 4.8, 5.2, 6.0, 6.8, 8.4 and 10.0
\AA. At nonlinear geometries, the grid included all 315 unique
combinations of $(r_{12}$, $r_{23}$ and $r_{31})$ taken from this
set that satisfy the triangle inequality, $r_{31}<r_{12}+r_{23}$.
We also added additional points for which $r_{12}$ and $r_{23}$
are in this set and $r_{31}$ takes values larger than 10.0 \AA\ in
steps of 1.6 \AA. This produced another 56 points, such as, for
example, $(6.0,6.0,11.6)$ \AA. A grid of 120 linear configurations
($r_{31}=r_{12}+r_{23}$) was formed by taking all possible
combinations of $r_{12}$ and $r_{23}$ from the above set with the
additional distance 5.6 \AA. We calculated electronic energies at
a total of 491 geometries.

Interaction energies for the $^4A'$ state were calculated with
respect to the 3-atom $S+S+S$ dissociation limit using RCCSD(T)
calculations with the s-contracted cc-pV5Z basis set described
above. All interaction energies were corrected for BSSE using the
counterpoise correction. \cite{Boys:1970}

\subsubsection{Fitting and interpolation}

Even with almost 500 points, the grid is fairly sparse and the
quality of interpolation is important. However, many interpolation
procedures in $N$ dimensions work only for a ``product" grid, in
which points are available at all combinations of a set of
coordinate values. This cannot be provided in pure bond length
coordinates because of the need to satisfy the triangle condition.
We therefore chose to use the interpolant-moving least squares
(IMLS) method. \cite{Ishida:2003} In this approach, approximate
gradients and Hessians are calculated at each grid point and
stored. The fitting function is represented as a weighted sum of
Taylor expansions of second order about the grid points.

The IMLS procedure has two parameters, the power $p$ that
determines the shape and range of the weight function and a
smoothing parameter $\epsilon$ that removes the flat spots at data
points. Figure 5 shows IMLS fits to the potential for $D_{\infty
h}$ geometries in the region of the conical intersection, with
$p=6$ and $\epsilon$ values of 0.03 and 0.05 \AA. The grid points
and the surface of Colavecchia {\em et al.}
\cite{Colavecchia:Li3:2003} are also shown for comparison. The
value of $\epsilon$ used here is large in comparison with previous
applications \cite{Ishida:2003, Ishida:1997, Ishida:1998,
Ischtwan:1994} and produces a root-mean-square error of 9.72
cm$^{-1}$. The {\it relative} error was larger than 1\% at 67
geometries, 21 of which lie below the 3-body dissociation limit
and are enclosed by 0.5 contour of the switching function (see
Eq.\ (\ref{switch_eq}) below). There are only 5 geometries at
which the relative error is larger than 10\%, and these lie either
near the conical intersection or near the zero of the potential
(where the relative error does not reflect the accuracy).

\begin{figure} [tbp]
\begin{center}
\includegraphics[width=60mm,angle=-90]{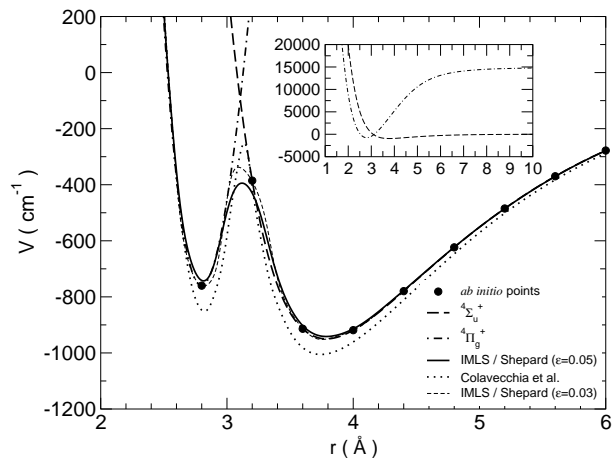}
\caption{Comparison of different fitted potentials with the {\em
ab initio} energies for the quartet ground state of Li$_3$ at
$D_{\infty h}$ geometries.} \label{fig5}
\end{center}
\end{figure}

\subsubsection{The long-range potential}

The long-range potential, $V_{\rm LR}({\bf r})$, was constructed
as a sum of pairwise potentials and a nonadditive part as in Eq.\
(\ref{Eqnonad}). The need to include the nonadditive term has been
described in detail in ref.\ \onlinecite{Cvitas:longrange:2006}.
In particular, the sum of pair potentials gives an {\em isotropic}
atom-diatom $C_6$ coefficient, whereas in reality there are
important anisotropic terms. For the pairwise potentials $V_{\rm
dimer}(r)$ we used the Li$_2$ RKR potential of Linton {\em et
al.}, \cite{Linton:1999} which extrapolates to a three-term
dispersion expression with coefficients $C_6$, $C_8$ and $C_{10}$
as given by Yan {\em et al.} \cite{Yan:1997} The nonadditive
potential was constructed as described in ref.\
\cite{Cvitas:longrange:2006} from the Axilrod-Teller-Muto
triple-dipole term, \cite{Axilrod:1943, Muto:1943} the third-order
dipole-dipole-quadrupole dispersion energy, \cite{Bell:1970,
Doran:1971} the fourth-order dipole dispersion energy,
\cite{Bade:1957, Bade:1958} and an atom-diatom term that is
short-range in one of the three distances,
\cite{Cvitas:longrange:2006}
\begin{eqnarray}
\nonumber V_{\rm 3,rep}(\mathbf{r})=
&-&[A+BP_2(\cos\theta)] \exp(-Cx) r_{23}^{-3}r_{31}^{-3} \nonumber\\
&\times& [D_6(r_{23}) D_6(r_{31})]^{1/2} \nonumber\\
&+& \hbox{cyclic permutations}, \label{expform}
\end{eqnarray}
where $x=(r_{12}-r_0)/r_0$ and $\theta$ is a Jacobi angle
approximated in terms of internal angles $\varphi_i$ as
\begin{equation}
P_2(\cos\theta)\simeq-\frac{1}{2}
(1+3\cos\varphi_1\cos\varphi_2\cos\varphi_3). \label{sym}
\end{equation}
All the non-additive terms were damped with a combination of
Tang-Toennies damping functions $D_6$ and $D_8$ \cite{Tang:1984}
as described in ref.\ \onlinecite{Cvitas:longrange:2006}. Three-body
dispersion coefficients $Z_{111}$ and $Z_{112}$ were taken from
refs.\ \onlinecite{Yan:1996} and \onlinecite{Patil:1997},
respectively, $Z_{1111}$ was estimated as described in
ref.~\onlinecite{Cvitas:longrange:2006}, and the parameters $A$,
$B$ and $C$ were determined as described in ref.\
\cite{Cvitas:longrange:2006} by fitting to the isotropic and
anisotropic dispersion coefficients for Li +
Li$_2$($^3\Sigma^{+}_u$) obtained as a function of $r$ by
R\'{e}rat {\em et al.} \cite{Rerat:2003}

The long-range form $V_{\rm LR}({\bf r})$ is designed to be valid
when {\it any} of the atom-atom distances is large. The
short-range (IMLS) and long-range potentials were joined using the
switching function
\begin{equation}
S({\bf r})=\frac{1}{2}\tanh[1-s_1(r_{12}+r_{21}+r_{31}-s_2)],
\label{switch_eq}
\end{equation}
with $s_1$ set to 0.7 \AA$^{-1}$ and $s_2$ to 20 \AA.

\begin{figure} [tbp]
\begin{center}
\includegraphics[width=80mm,angle=-90]{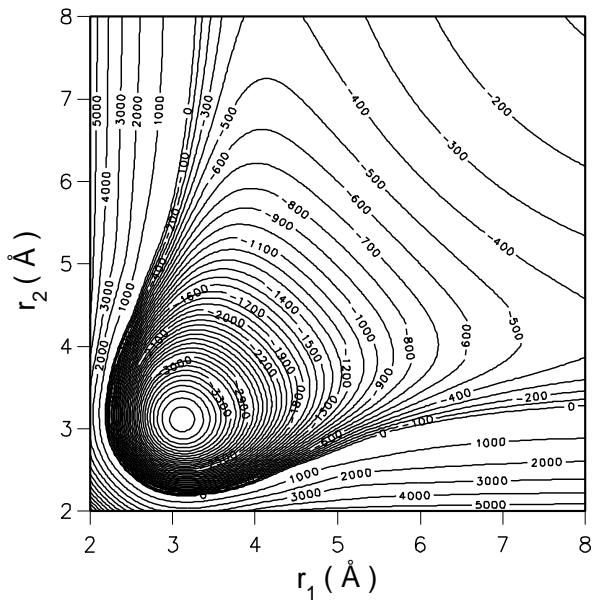}
\caption{The fitted quartet ground-state potential of lithium
trimer for a bond angle of $60^\circ$. Contours are labeled in
cm$^{-1}$.} \label{fig6}
\end{center}
\end{figure}

\begin{figure} [tbp]
\begin{center}
\includegraphics[width=80mm,angle=-90]{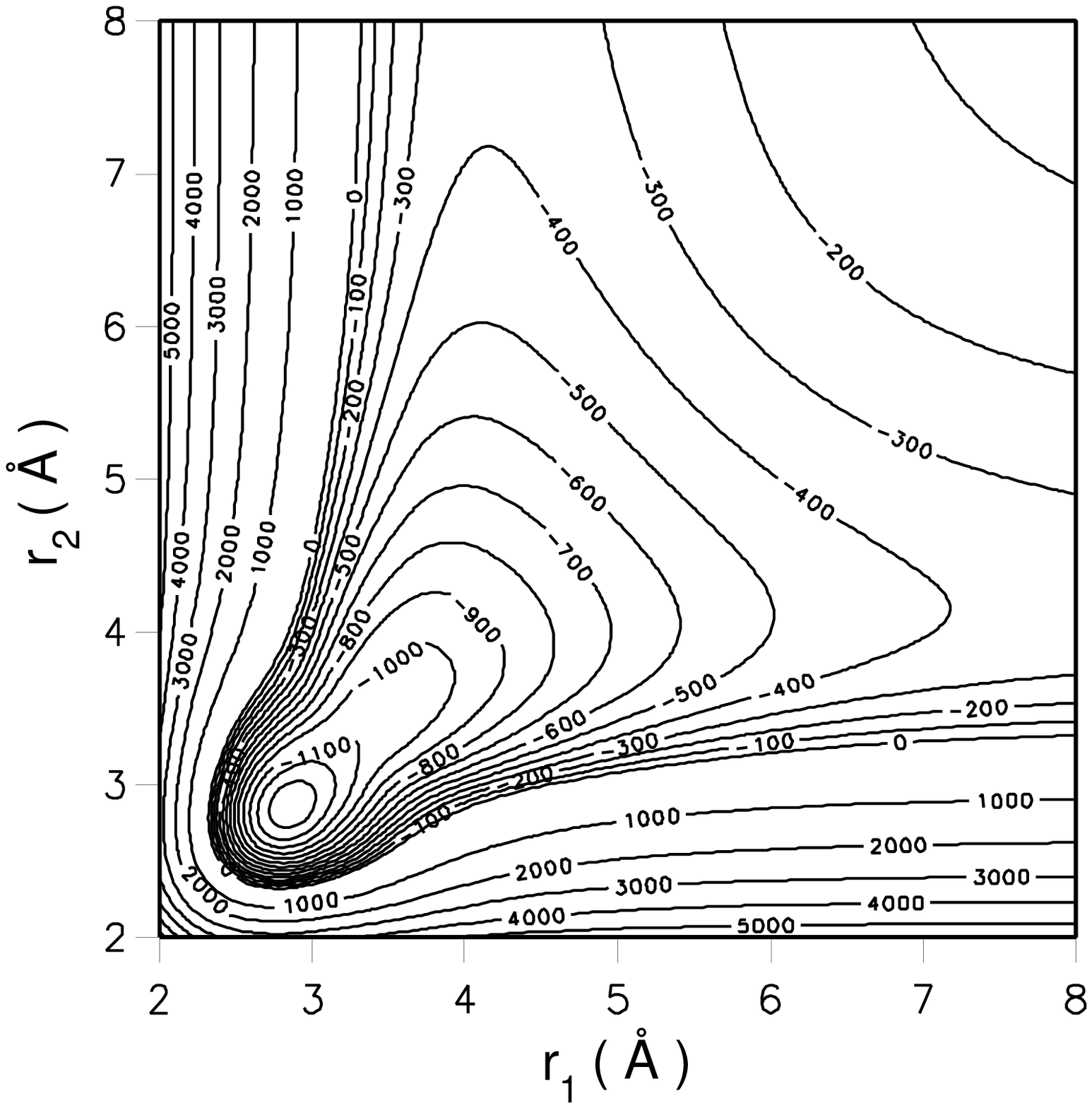}
\caption{The fitted quartet ground-state potential of lithium
trimer for a bond angle of $120^\circ$. Contours are labeled in
cm$^{-1}$.} \label{fig7}
\end{center}
\end{figure}

\begin{figure} [tbp]
\begin{center}
\includegraphics[width=80mm,angle=-90]{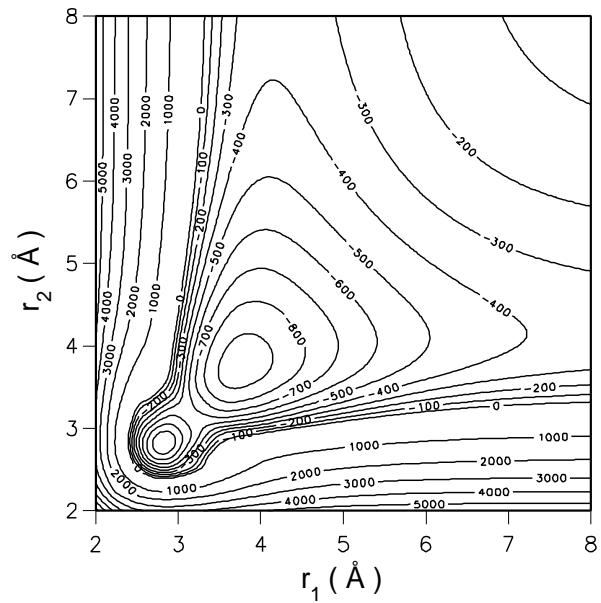}
\caption{The fitted quartet ground-state potential of lithium
trimer for a bond angle of $150^\circ$. Contours are labeled in
cm$^{-1}$.} \label{fig8}
\end{center}
\end{figure}

Contour plots of the final potential for three different interbond
angles $\alpha$ are shown in Figures \ref{fig6} to \ref{fig8}.

\section{Li + \protect{Li$_2$} scattering calculations}

\subsection {Methodology}

Both linear and equilateral configurations of the Li$_3$ collision
complex (and all geometries in between) lie lower in energy than
Li + Li$_2$. The potential energy surface thus allows barrierless
atom exchange (rearrangement) collisions even at limitingly low
collision energies. Calculations using an inelastic formalism
(including only one arrangement channel) are therefore inadequate.
Accordingly, we have carried out quantum reactive scattering
calculations to obtain collision cross sections. The methodology
has been described in our previous work on ultracold collisions in
Na + Na$_2$ \cite{Soldan:2002,Quemener:2004} and K + K$_2$
\cite{Quemener:2005} and also in studies of insertion reactions
such as $\mbox{N}(^2D)+\mbox{H}_2 \rightarrow \mbox{NH}+\mbox{H}$,
\cite{Honvault:1999} $\mbox{O}(^1D)+\mbox{H}_2 \rightarrow
\mbox{OH}+\mbox{H}$ \cite{Honvault:2001, Aoiz:2001} and
$\mbox{S}(^2D)+\mbox{H}_2 \rightarrow \mbox{SH}+\mbox{H}$
\cite{Honvault:2003} at thermal energies.

The time-independent Schr\"{o}dinger equation is solved using a
coupled-channel method in body-fixed democratic hyperspherical
coordinates \cite{Launay:1989} using a diabatic-by-sector method.
In each sector, the wavefunction is expanded in the eigenstates of
a reference hamiltonian at a fixed hyperradius. The reference
hamiltonian includes potential energy terms and kinetic energy
terms arising from deformation and rotation around the axis of
least moment of inertia. Its eigenfunctions are obtained by
variational expansion in a set of pseudohyperspherical harmonics.
The resulting coupled equations are integrated using the
log-derivative method of Manolopoulos. \cite{Manolopoulos:1986} At
the point where interchannel couplings are small enough to be
neglected, the coupled-channel solutions are matched onto
atom-diatom functions propagated inwards in Jacobi coordinates to
take account of the isotropic long-range and centrifugal
potentials.

The diatom wavefunctions for Hund's case (b), which is appropriate
for Li$_2$ $(^3\Sigma_{u}^{+})$, are labeled with a vibrational
quantum number $v$ and a mechanical rotational quantum number $n$;
$n$ couples with the diatomic electron spin $s=1$ to give a
resultant $j$. However, for Li$_2$ the splitting between states of
the same $n$ but different $j$ are very small. We therefore do not
introduce the spin explicitly and $n$ and $j$ are equivalent in
our calculations.

The matching procedure yields the reactance matrix $K$, the
scattering matrix $S$ and the transition matrix $T=I-S$, where $I$
is a unit matrix. The differential cross sections, averaged over
initial $m_n$ and summed over final $m'_n$ states, are given by
\begin{eqnarray}
\left(\frac{\mbox{d}\sigma}{\mbox{d}\Omega}\right)_{\tau vn
\rightarrow \tau' v'n'} &&= \label{diffcross}
\\
\nonumber \frac{1}{4(2n+1)k_{\tau v n}^2}&& \left| \sum_{J K K'}
(2J+1)T^J_{\tau' v' n' K' \tau v n K}
d^J_{K K'}(\vartheta)\right|^2,
\end{eqnarray}
where $\tau$ labels the arrangement, $v$ and $n$ indicate the
vibrational and rotational quantum numbers of the diatom, $K$
labels the $z$ component of $J$ and $n$ in the body-fixed frame,
and $k_{\tau vn}$ is the corresponding wave vector. $\Omega$ is an
element of solid angle at a scattering angle $\vartheta$ with
respect to the initial approach direction in the centre-of-mass
frame and the $d$ functions are reduced Wigner rotation matrices.

Integral state-to-state cross sections are obtained by integrating
Eq.\ \ref{diffcross},
\begin{equation}
\sigma_{\tau vn \rightarrow \tau' v'n'}= \frac{\pi}{(2n+1)k_{\tau
v n}^2} \sum_{J K K'} (2J+1) \left| T^J_{\tau' v' n'
K' \tau v n K}\right|^2. \label{intcross}
\end{equation}
At ultralow energies, only $T_{ii}$ matrix elements with
atom-diatom end-over-end angular momentum $l=0$ contribute
significantly to the cross sections. The elastic and total
inelastic cross sections in terms of $T_{ii}$ are then
\begin{eqnarray}
\sigma_{\rm elas}&=&\frac{\pi}{k^2}|T_{ii}|^2,
\label{sig_elas} \\
\sigma_{\rm inel}^{\rm tot}&=& \frac{\pi}{k^2}\left[
1-|1-T_{ii}|^2 \right], \label{sig_inel}
\end{eqnarray}
where the $\tau vn$ subscript has been dropped to simplify
notation. At low energy, $S_{ii}$ and $T_{ii}$ are conveniently
parameterized in terms of an energy-dependent complex scattering
length $a(k)=\alpha(k)-{\rm i}\beta(k)$, where
\begin{equation}
a(k)=\frac{1}{{\rm i}k}\left(\frac{1-S_{ii}}{1+S_{ii}}\right)
=\frac{1}{{\rm i}k}\left(\frac{T_{ii}}{2-T_{ii}}\right).
\label{aexact}
\end{equation}
The cross sections can be written exactly in terms of the
energy-dependent scattering length,
\begin{eqnarray}
\sigma_{\rm elas}&=&\frac{4\pi|a|^2}{1+k^2|a|^2+2k\beta},
\label{sig_elas_sl} \\
\sigma_{\rm inel}^{\rm tot}
&=&\frac{4\pi\beta}{k(1+k^2|a|^2+2k\beta)}. \label{sig_inel_sl}
\end{eqnarray}
The scattering length becomes constant at limitingly low energy.
If $ka\ll 1$, which is true at limitingly low energy except near a
zero-energy resonance, Eq.\ \ref{aexact} reduces to the expression
commonly used for the zero-energy scattering length,
\cite{Balakrishnan:scat-len:1997}
\begin{equation}
a=\frac{1}{2{\rm i}}\lim_{k\rightarrow 0} \frac{T_{ii}}{k}.
\label{scatlen_extr}
\end{equation}

Elastic and inelastic cross sections at ultralow energies are
governed by the Wigner threshold laws, \cite{Wigner:1948}
\begin{eqnarray}
\nonumber \sigma_{{\rm elas}}&\sim&E^{2l}, \\
\sigma_{{\rm inel}}&\sim&E^{l-1/2},
\label{wig_laws}
\end{eqnarray}
where $E$ is the collision energy and $l$ is the end-over-end
angular momentum. In the presence of a long-range potential $-C_s
r^{-s}$, elastic cross sections have an additional term $\sim
E^{s-3}$ that dominates for higher $l$ at ultralow energies.

\subsection{Computational details}

We have performed coupled-channel atom-diatom collision
calculations on all possible mixtures of the two isotopes of
lithium, $^7$Li and $^6$Li. The $^7$Li nucleus is a fermion with
nuclear spin $3/2$, so that the atom, including its electrons,
behaves as a composite boson in cold dilute gases. The $^6$Li
nucleus is a boson with nuclear spin 1, while the atom is a
composite fermion. The present calculations are restricted to
collisions in which all three of the atoms are in spin-stretched
states (with $F=I+S=F_{{\rm max}}$ and $|M_F|=F$). For these
states the nuclear spin wavefunction is symmetric with respect to
any exchange of nuclei. However, the electronic wavefunction for
the quartet state, which depends parametrically on nuclear
coordinates, is antisymmetric with respect to exchange of two
identical nuclei. This means that the spatial nuclear wavefunction
must be symmetric under exchange of any two fermionic nuclei
(bosonic alkali metal atoms) and antisymmetric under exchange of
any two bosonic nuclei (fermionic alkali metal atoms). The
symmetry under exchange of identical nuclei is easy to impose in
our program by selecting pseudohyperspherical harmonics of
appropriate symmetry in the basis set.

Propagation of the coupled-channel solutions is performed
separately for each partial wave $J^{\Pi}$, labeled by the
spin-free total angular momentum $J$, and parity eigenvalue $\Pi$.
The parity is $(-1)^{n+l}$, and only $l$ values that satisfy the
triangle inequality, $|J-n| \leq l \leq J+n$, are permitted.

The basis set for $^7$Li$_3$ included all eigenstates of the
reference hamiltonian (see above) that match onto $^7$Li$_2$
states below the $v=7$ manifold. Only even values of $n$ are
allowed for $^7$Li$_2$ in spin-stretched states. The resulting
number of coupled equations varied between 97, for total angular
momentum $J=0$, and 827, for $J=10$. For the $^6$Li$_3$ system,
including only odd-$n$ states of $^6$Li$_2$ below the $v=7$
manifold gave a basis set that varied in size between 85 for $J=0$
and 782 for $J=11^{-}$. For $^6$Li$^7$Li$_2$ and $^7$Li$^6$Li$_2$,
where the symmetry is reduced, the basis set even for $J=0$
consisted of 272 and 263 functions, respectively. The calculation
of eigenstates of the reference hamiltonian involved
diagonalizations of matrices of sizes between 1240 and 2136 for
$^7$Li$_3$, between 1220 and 2162 for $^6$Li$_3$, and between 3660
and 6488 for the isotopically mixed systems.

The coupled equations were integrated across a range of
hyperradius $\rho$ between $5$ $a_0$ and 45 $a_0$ with a
log-derivative sector width of 0.1 $a_0$. Within each sector, the
wavefunction is propagated using the log-derivative method of
Manolopoulos, \cite{Manolopoulos:1986} with a step size keyed to
the local de Broglie wavelength. The propagation in the outer
region in Jacobi coordinates extended as far as $R=10000$ $a_0$
for collision energies of $E=1$ nK.

\subsection{Ultracold collisions}

\begin{figure} [tbp]
\begin{center}
\includegraphics[width=60mm,angle=-90]{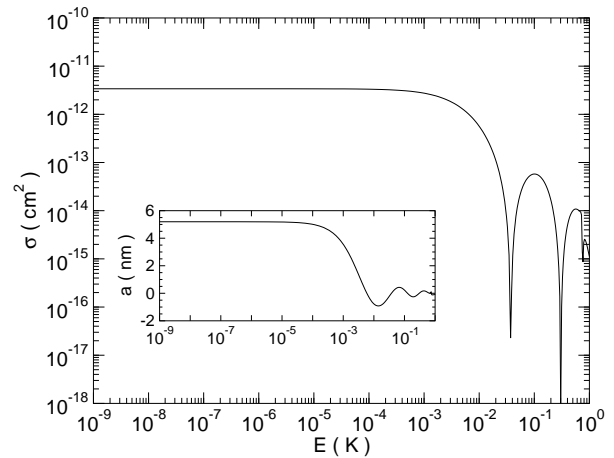}
\caption{Energy dependence of elastic cross sections for
$^7\mbox{Li}+\mbox{}^7\mbox{Li}_2(v_i=0, n_i=0)$. The
energy-dependent scattering length is shown in the inset.}
\label{fig9}
\end{center}
\end{figure}

\begin{figure} [tbp]
\begin{center}
\includegraphics[width=60mm,angle=-90]{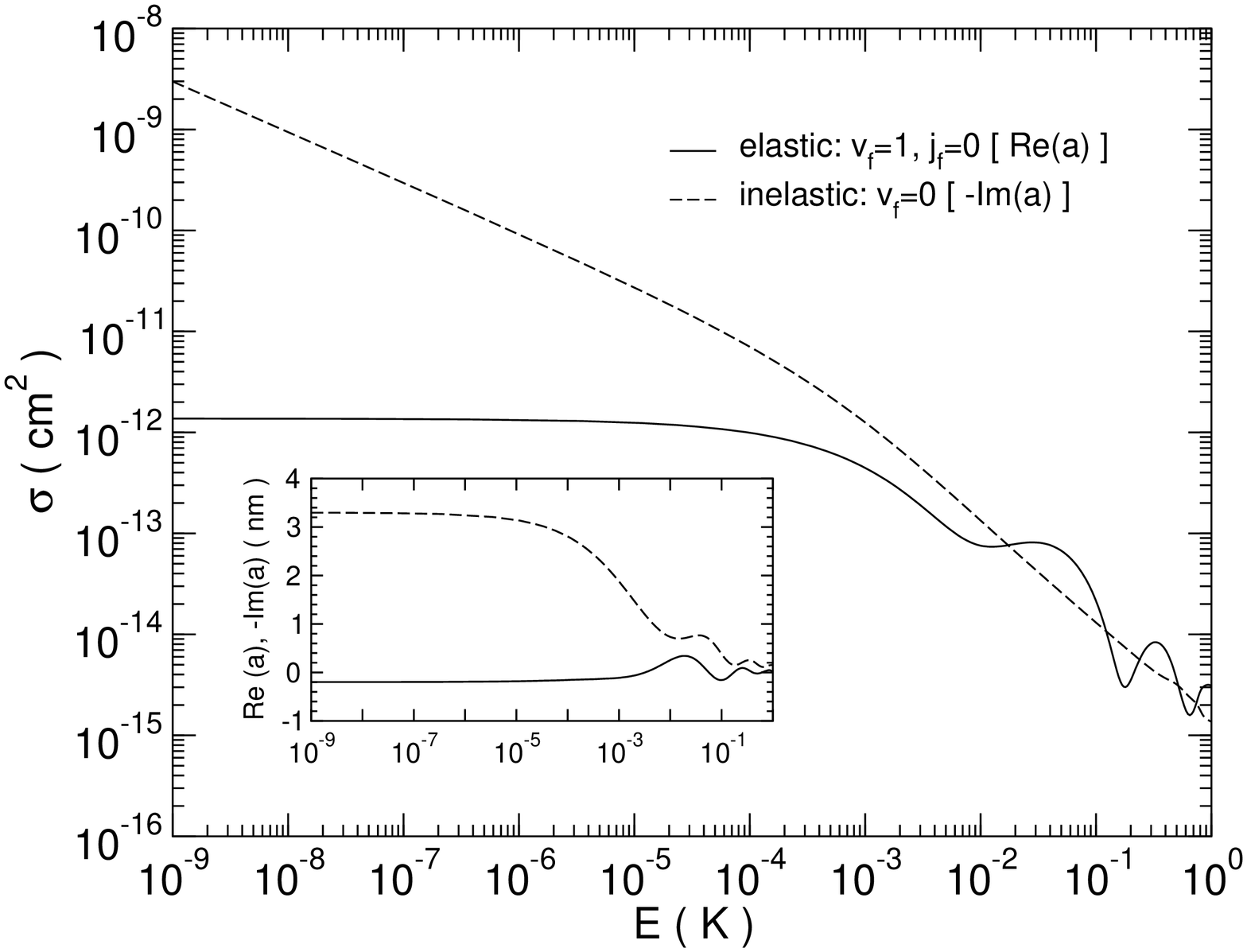}
\caption{Energy dependence of elastic and inelastic cross sections
for $^7\mbox{Li}+\mbox{}^7\mbox{Li}_2(v_i=1, n_i=0)$. The complex
energy-dependent scattering length is shown in the inset.}
\label{fig10}
\end{center}
\end{figure}

In the ultracold regime, the dominant contribution to cross
sections comes from the partial wave that includes the $l=0$
channel (s-wave scattering). For a particular initial
rovibrational state $(v,n)$, the cross section is thus calculated
from the partial wave labeled by the total angular momentum $J=n$
and parity $\Pi=(-1)^n$. Figures \ref{fig9} and \ref{fig10} show
the $J=0$ cross sections for $^7$Li + $^7$Li$_2$ for initial
states $(v,n)=(0,0)$ and $(1,0)$. The insets show the real and
imaginary part of the complex scattering length, calculated using
Eq.\ (\ref{aexact}).

\begin{figure} [tbp]
\begin{center}
\includegraphics[width=60mm,angle=-90]{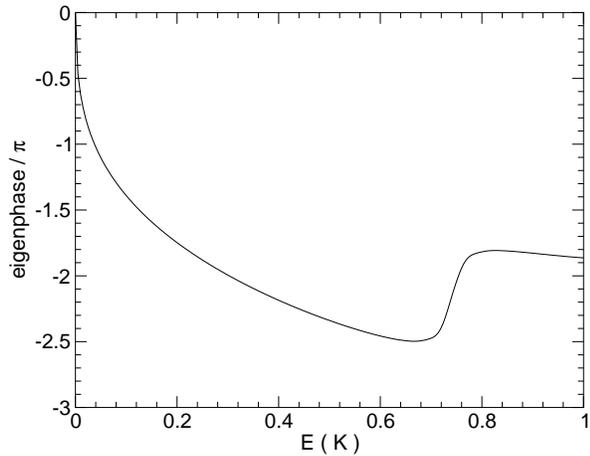}
\caption{Energy dependence of the phase shift $\delta$ for
$^7\mbox{Li}+\mbox{}^7\mbox{Li}_2(v_i=0, n_i=0)$.}
\label{fig11}
\end{center}
\end{figure}

The energy dependence is particularly simple for collisions
involving a ground-state $^7$Li$_2$ molecule, $(v,n)=(0,0)$. Only
elastic scattering is allowed below 2.3 K (the energy of the
$(0,2)$ threshold). The elastic cross sections are essentially
constant below $\sim 10^{-5}$ K, and oscillate at higher energies.
The s-wave cross sections exhibit sudden drops to zero at 37 mK
and 300 mK and a feature around 750 mK. The one-channel scattering
is characterized by a phase shift $\delta$ that can be extracted
from the $1\times 1$ $K$ matrix element, $K_{11}=\tan \delta$.
\cite{Ashton:1983} The energy dependence of the phase shift is
shown in Figure \ref{fig11}, and the zeroes in the cross section
can be associated with $\delta$ passing through a multiple of
$\pi$. The profile of the cross section follows $\sim
\sin^2\delta$. However, it should be noted that partial waves with
$l>0$ make significant contributions at collision energies above
about 1 mK, so that the zeroes in the s-wave cross section will be
washed out in the total elastic cross section.

The feature at $\sim 750$ mK is different. It is associated with a
sharp rise in $\delta$ through $\pi$, superimposed on the falling
background. This is characteristic of a scattering resonance, and
in this case is due to a Feshbach resonance involving a quasibound
state associated with a rotationally excited threshold.

\begin{figure} [tbp]
\begin{center}
\includegraphics[width=60mm,angle=-90]{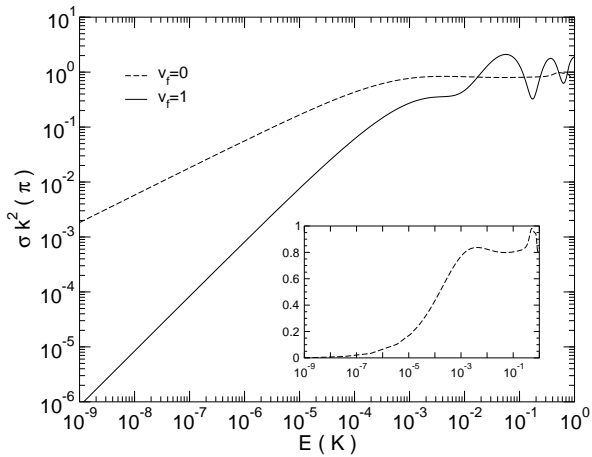}
\caption{Energy dependence of elastic and sum of inelastic
probabilities, $\sigma k^2/\pi$, for
$^7\mbox{Li}+\mbox{}^7\mbox{Li}_2(v_i=1,n_i=0)$.} \label{fig12}
\end{center}
\end{figure}

\begin{figure} [tbp]
\begin{center}
\includegraphics[width=75mm,angle=0]{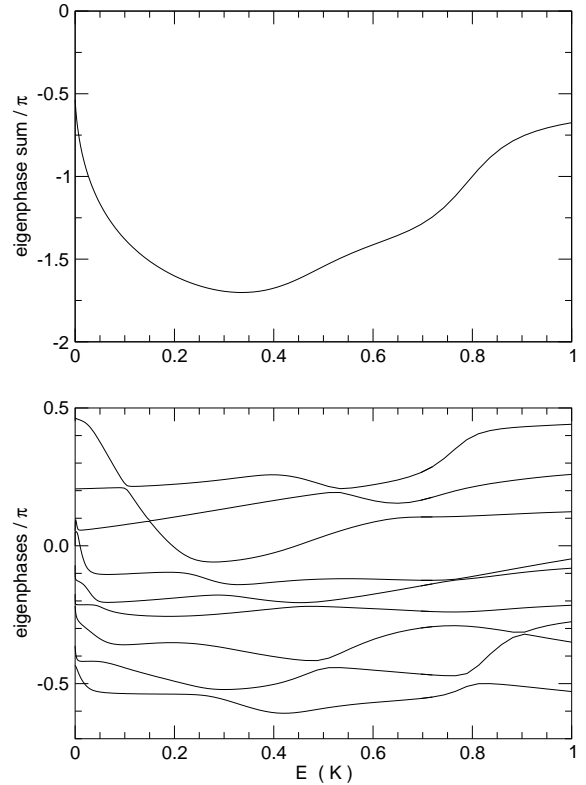}
\caption{Eigenphase sum (upper panel) and individual eigenphases
(lower panel) for
$^7\mbox{Li}+\mbox{}^7\mbox{Li}_2(v_i=1,n_i=0)$.} \label{fig13}
\end{center}
\end{figure}

For the vibrationally excited initial state $(v,n)=(1,0)$ the
above arguments generalize. The S-matrix is now an $9\times 9$
matrix with 8 inelastic (or reactive) channels corresponding to
$v=0$, $n=0, 2, \ldots 14$. The elastic cross sections, Figure
\ref{fig12}, still show dips, but these do not reach zero because
the elastic S-matrix element cannot be 1 in the presence of
inelastic scattering. The quantity that now shows a rise of $\pi$
across a resonance is the eigenphase sum, which is the sum of
phases obtained from eigenvalues of the $K$ (or $S$) matrix. The
eigenphase sum and the individual eigenphases are shown in Figure
\ref{fig13}. It may be seen that there are two broad resonances
centered at about 0.5 K and 0.8 K, but that the phase change
associated with them is spread between all the channels. The
individual eigenphases show a complicated pattern with many
avoided crossings.

The inelastic cross sections show a slope of $-1/2$ on a log-log
scale at ultralow temperatures, as predicted by the Wigner
threshold law, Eq.\ (\ref{wig_laws}). Above about $10^{-2}$ K, the
slope changes to $-1$. The total inelastic probability, shown in
Figure \ref{fig12}, increases with collision energy according to
the Wigner laws and saturates close to unity outside the Wigner
region. \cite{Soldan:2002} At this point the energy dependence of
the cross sections is governed by the kinematic prefactor $1/k^2$
in Eq.\ (\ref{sig_inel}). A high inelastic probability is a
feature of reactions proceeding over a deep well.
\cite{Honvault:1999, Honvault:2001, Aoiz:2001, Soldan:2002,
Honvault:2003, Quemener:2005}

\begin{figure} [tbp]
\begin{center}
\includegraphics[width=60mm,angle=-90]{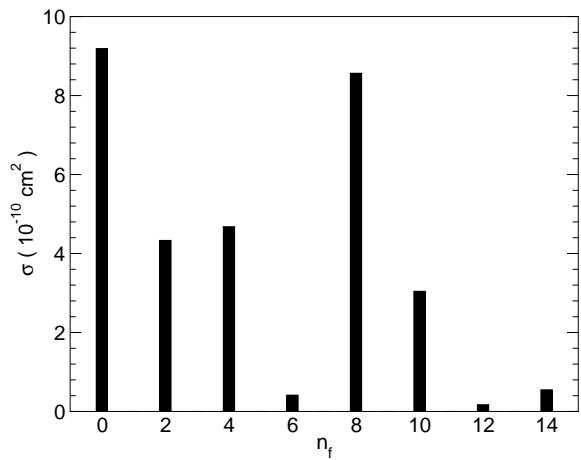}
\caption{Final rotational distributions for
$^7\mbox{Li}+\mbox{}^7\mbox{Li}_2(v_i=1,n_i=0)$ at 0.928 nK.}
\label{fig14}
\end{center}
\end{figure}

The final state rotational distributions at $\sim 1$ nK are shown
in Figure \ref{fig14}. The oscillatory structure is similar to
that found earlier in ultracold collisions of Na + Na$_2$
\cite{Soldan:2002} and similar structures have been observed in
vibrational predissociation of Van der Waals complexes. The
oscillation is thought to be due to a rotational rainbow effect.
\cite{Hutson:ArH2:1983} In classical terms, the angular momentum
imparted to the Li$_2$ molecule is zero if the kinetic energy is
released at linear or T-shaped geometries, but large at around
$\theta=45^\circ$. In this model, the oscillations arise from
interference between classical trajectories on either side of the
maximum.

The relative rotational distributions do not change significantly
across the Wigner regime, although individual cross sections vary
over five orders of magnitude between 1 nK and 100 mK because of
the $k^{-1}$ dependence of the Wigner threshold laws.

We have studied collisions involving a range of different initial
molecular states. Results at a collision energy of $\sim 1$ nK are
presented in Table \ref{crossrate_exc}. There is no {\em
systematic} dependence of the cross sections on the initial
molecular state. Instead, the cross sections show essentially
random scatter about a mean value. The origin of this behavior is
discussed below. At 1 nK, the inelastic cross sections are
typically three orders of magnitude higher than the elastic cross
sections. The ratio decreases with the collision energy according
to the Wigner laws, Eq.\ (\ref{wig_laws}), until it reaches $\sim
1$ in the millikelvin range. Sympathetic and evaporative cooling
of atoms and molecules depend on a low value ($< 10^{-2}$) of this
ratio, so it is unlikely that such cooling mechanisms will be
possible for alkali metal dimers in low-lying vibrationally
excited states.

It has previously been found that collisions involving
rotationally excited diatoms can cause unusually efficient and
specific energy transfer when the collision time is longer than
the rotational period and there is a match between the rotational
and vibrational frequencies. \cite{Stewart:1988, Magill:1988} This
has been termed quasiresonant energy transfer to distinguish it
from energy-resonant behaviour arising from energy matching
between the initial and final states. Quasiresonant energy
transfer has been explained in terms of adiabatic invariance
theory \cite{Hoving:1989} with the assumption that the coupling
between the degrees of freedom is not strong. Such behaviour has
also been studied at ultralow temperatures. \cite{Forrey:qr:1999,
Forrey:2002, Florian:2004, Tilford:2004} Inelastic transitions in
some highly excited rotational levels of hydrogen molecules were
found to be dramatically suppressed when quasiresonant transitions
were energetically forbidden. However, the highest initial
rotational states considered for $^7$Li$_2$ have $j=10$, for which
the rotational energy is only about half the vibrational spacing.
We thus do not expect to observe quasiresonant enhancements of the
type described in refs.\ \onlinecite{Stewart:1988} and
\onlinecite{Magill:1988}. In addition, we observe no near-resonant
enhancement of the inelastic rate for initial state $v=2$, $j=0$,
for which a transition to the $v=0$, $j=20$ level is exothermic by
only 3.33 K. As may be seen in Table \ref{crossrate_exc}, the
inelastic cross section from this level is actually the {\it
lowest} of all those calculated.

The atom-molecule inelastic loss rates observed in experiments on
ultracold boson dimers ($^{133}$Cs$_2$), \cite{Staanum:2006,
Zahzam:2006} are around $10^{-10}$ cm$^3$ s$^{-1}$, which is
consistent with the cross sections obtained here for the bosonic
lithium dimer ($^7$Li$_2$).

\begin{table}[tbp]
\begin{center}
\begin{tabular}{cccc} \hline
$v_i$, $n_i$ \quad&\quad $\sigma_{{\rm elas}}$ (cm$^{2}$)\quad
&\quad $\sigma_{{\rm inel}}$ (cm$^{2}$) \quad &
$\sigma_{{\rm inel}}/\sigma_{{\rm elas}}$ \\
\hline
0, 0  & $3.39 \times 10^{-12}$ & $-$ & $-$ \\
0, 2  & $4.87 \times 10^{-12}$ & $6.56 \times 10^{-10}$ & 135 \\
0, 4  & $3.90 \times 10^{-13}$ & $9.55 \times 10^{-10}$ & 2450 \\
0, 6  & $7.72 \times 10^{-13}$ & $9.42 \times 10^{-10}$ & 1220 \\
0, 8  & $1.57 \times 10^{-12}$ & $2.04 \times 10^{-9}$  & 1300 \\
0, 10 & $9.26 \times 10^{-13}$ & $2.55 \times 10^{-9}$  & 2750 \\
\hline
1, 0  & $1.37 \times 10^{-12}$ & $3.09 \times 10^{-9}$ & 2260 \\
1, 2  & $2.05 \times 10^{-12}$ & $3.00 \times 10^{-9}$ & 1460 \\
1, 4  & $8.00 \times 10^{-13}$ & $1.14 \times 10^{-9}$ & 1425 \\
1, 6  & $8.46 \times 10^{-13}$ & $1.43 \times 10^{-9}$ & 1690 \\
1, 8  & $1.74 \times 10^{-12}$ & $1.96 \times 10^{-9}$ & 1130 \\
1, 10 & $1.38 \times 10^{-12}$ & $1.53 \times 10^{-9}$ & 1110 \\
\hline
2, 0  & $5.17 \times 10^{-13}$ & $4.77 \times 10^{-10}$ & 920 \\
2, 2  & $1.02 \times 10^{-12}$ & $1.96 \times 10^{-9}$  & 1920 \\
2, 4  & $1.25 \times 10^{-12}$ & $1.56 \times 10^{-9}$  & 1250 \\
2, 6  & $8.83 \times 10^{-13}$ & $1.48 \times 10^{-9}$  & 1680 \\
2, 8  & $9.85 \times 10^{-13}$ & $1.87 \times 10^{-9}$  & 1900 \\
2, 10 & $1.32 \times 10^{-12}$ & $1.95 \times 10^{-9}$  & 1480 \\
\hline
3, 0  & $9.29 \times 10^{-13}$ & $8.57 \times 10^{-10}$ & 920 \\
3, 2  & $1.06 \times 10^{-12}$ & $1.43 \times 10^{-9}$  & 1350 \\
3, 4  & $1.16 \times 10^{-12}$ & $2.42 \times 10^{-9}$  & 2090 \\
3, 6  & $1.77 \times 10^{-12}$ & $2.61 \times 10^{-9}$  & 1470 \\
3, 8  & $2.55 \times 10^{-12}$ & $3.85 \times 10^{-9}$  & 1510 \\
3, 10 & $1.27 \times 10^{-12}$ & $1.96 \times 10^{-9}$  & 1540 \\
\cline{1-4}
\end{tabular}
\vspace{0.5cm} \caption{Elastic and total inelastic cross sections
and rate coefficients for $^7\mbox{Li} +
\mbox{}^7\mbox{Li}_2(v_i,n_i)$ at a collision energy of 0.928 nK
for different initial states of the molecule.}
\label{crossrate_exc}
\end{center}
\end{table}

\subsection{Cold collisions}

\begin{figure} [tbp]
\begin{center}
\includegraphics[width=60mm,angle=-90]{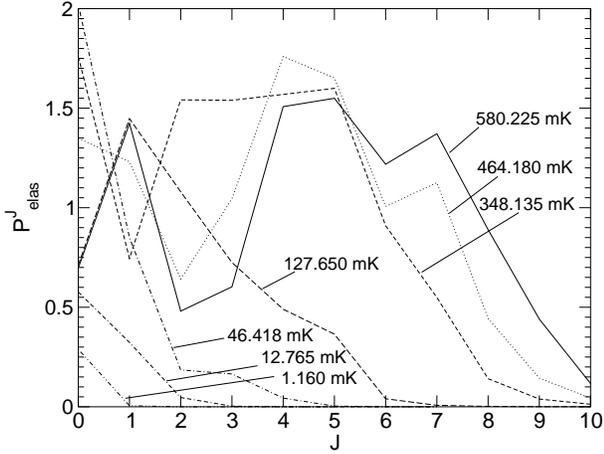}
\caption{Elastic probabilities, $\sigma k^2 / \pi$, as a function
of total angular momentum {\em J} for
$^7\mbox{Li}+\mbox{}^7\mbox{Li}_2(v=1,n=0)$.} \label{fig15}
\end{center}
\end{figure}

\begin{figure} [tbp]
\begin{center}
\includegraphics[width=60mm,angle=-90]{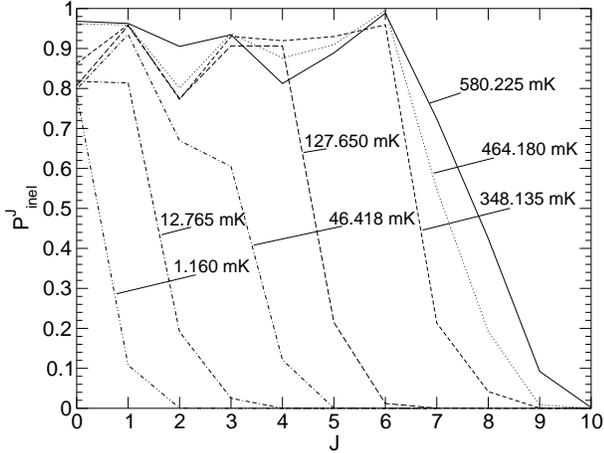}
\caption{Inelastic probabilities, $\sigma k^2 / \pi$, as a
function of total angular momentum {\em J} for
$^7\mbox{Li}+\mbox{}^7\mbox{Li}_2(v=1,n=0)$.} \label{fig16}
\end{center}
\end{figure}

At collision energies above about 1 mK, partial waves with
end-over-end angular momentum $l>0$ contribute significantly to
the overall cross sections. The convergence of the elastic and
inelastic probabilities $\sigma k^2 / \pi$ with total angular
momentum $J$ is shown in Figures \ref{fig15} and \ref{fig16} for
collisions with molecules in the $(1,0)$ initial state. Partial
waves $J=0$ to 10 suffice to converge cross sections up to $\sim
580$ mK, where the contribution of $J=10$ to elastic and inelastic
cross sections is 2.35\% and 0.0655\%, respectively. It may be
seen that, for each energy, the inelastic cross sections are close
to 1 for partial waves up to an energy-dependent cutoff value. The
corresponding elastic cross sections oscillate around 1 until the
cutoff is reached. Above 100 mK, convergence is slower for elastic
probabilities, while below that it is slower for inelastic
probabilities.

The cross sections for partial waves with $J>0$ also show a simple
energy dependence. In the Wigner regime the inelastic cross
sections increase with slope $l-1/2$ on a log-log scale, with
$l=J$ for $n_i=0$. At energies above the centrifugal barrier they
decrease with slope $-1$.

\begin{figure} [tbp]
\begin{center}
\includegraphics[width=60mm,angle=-90]{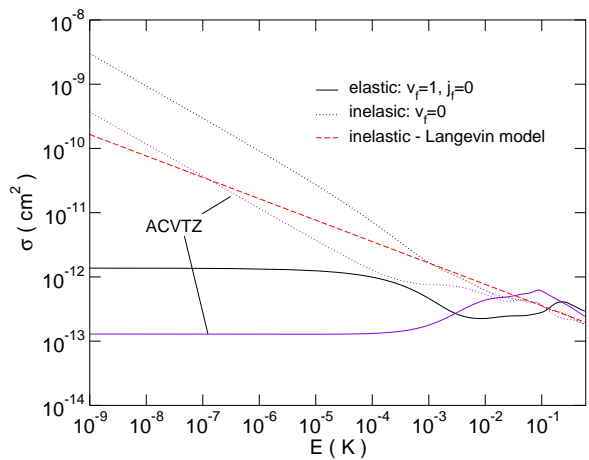}
\caption{[Color online] Elastic and total inelastic cross sections
for $^7\mbox{Li}+\mbox{}^7\mbox{Li}_2(v_i=1,n_i=0)$ on the
uncontracted cc-pV5Z and aug-cc-pCVTZ basis set potentials and the
inelastic cross sections in the Langevin model.} \label{fig17}
\end{center}
\end{figure}

When several partial waves contribute significantly to the cross
sections, the inelastic cross sections can be described
semiquantitatively by classical Langevin capture theory. The
assumption underlying this model is that the only condition for an
inelastic event to take place is for the projectile to surmount
the barrier of the effective potential resulting from the
repulsive centrifugal and attractive long-range potential. For a
long-range interaction of the form $-C_s/R^{-s}$ the effective
potential is
\begin{equation}
V^l_{\rm eff}(R) = \frac{\hbar^2 l(l+1)}{2 \mu R^2} - \frac{C_s}{R^s}.
\end{equation}
This has a barrier height of
\begin{equation}
V^l_{\rm max}=\left[\frac{\hbar^2 l(l+1)}{\mu} \right]^{s/(s-2)}
(sC_s)^{2/(2-s)} \left( \frac{1}{2} - \frac{1}{s} \right)
\label{c_barr1}
\end{equation}
at an atom-diatom separation of
\begin{equation}
R^l_{{\rm max}}=\left( \frac{s \mu C_s} {\hbar^2 l(l+1)}
\right)^{1/(s-2)}. \label{c_barr2}
\end{equation}
Taking $s=6$, $C_6=3085.54\ E_h a_0^6$, and $\mu=2 m_{{\rm
Li}}/3$, we obtain barrier heights of 2.78 mK, 14.4 mK, 40.8 mK,
87.9 mK, 161 mK for $l=1$ to 5, at atom-molecule separations
ranging from 94.25 $a_0$ for $l=1$ to 47.89 $a_0$ for $l=5$. At
these distances the interaction is dominated by dispersion and is
nearly isotropic. The Langevin expression for the inelastic cross
section is
\begin{equation} \sigma_{{\rm inel}}(E)=\frac{s\pi}{2} \left(
\frac{2}{s-2}\right)^{(s-2)/s} \left( \frac{C_s}{E} \right)^{2/s}.
\label{cross_langevin}
\end{equation}
The total inelastic cross sections are compared with this result
in Figure \ref{fig17} and it may be seen that there is
increasingly good agreement between the full calculations and the
Langevin model above about 30 mK.

To test that the agreement between the Langevin model and our
results is not fortuitous, we recalculated the cross sections
using partial waves $J=0$ to 10 using a different potential
surface, designated ACVTZ. The ACVTZ potential was obtained in the
same manner as described above, but with electronic energies
calculated using a considerably smaller basis set (aug-cc-pCVTZ).
The potential is significantly different, with a global $D_{3h}$
minimum 3873.37 cm$^{-1}$ deep at $r=3.125$ \AA\ and a $D_{\infty
h}$ minimum 930.29 cm$^{-1}$ deep at $r=3.780$ \AA. The two
potentials have the same long-range form, joined using the
switching function of Eq.\ (\ref{switch_eq}). The elastic and
inelastic cross sections calculated using the ACVTZ potential are
included in Figure \ref{fig17}. Once again the agreement with the
Langevin model is excellent above $\sim 30$ mK, though the two
potentials give cross sections that differ by about an order of
magnitude in the Wigner regime. We found similar agreement with
the model for collisions involving other initial molecular states.
The Langevin cross sections are independent of the initial state,
particle masses and the details of short-range potential.

Elastic cross sections are comparable to inelastic ones in the
millikelvin region and can be as high as twice the inelastic ones.
However, the ratio is never high enough to be favorable for the
prospects of evaporative or sympathetic cooling. Whenever the
Langevin model provides good estimates of inelastic cross
sections, it implies that the inelastic probabilities are close to
1, which in turn implies that elastic cross sections (Eq.\
(\ref{sig_elas})) and inelastic cross sections (Eq.\
(\ref{sig_inel})) are not drastically different.

The reaction $^7\mbox{Li}+\mbox{}^7\mbox{Li}_2$  has no barrier
for either linear or perpendicular approaches. The deep well in
the potential energy surface suggests that the reaction proceeds
via an insertion mechanism involving complex formation. It is
therefore interesting to compare the scattering results obtained
here with those for insertion reactions that have been studied at
ordinary temperatures ($\sim 100$ meV). \cite{Honvault:1999,
Honvault:2001, Aoiz:2001, Honvault:2003} These reactions are also
characterized by high inelastic and reactive probabilities. The
final-state vibrational distributions decrease with $v$ and the
rotational distributions peak at high $n$ for each final $v$. If
all product states are equally probable, the product state
distribution is proportional to the density of available states,
which depends on the rotational degeneracy and the density of
translational states. The resulting differential cross sections
display forward-backward symmetry. This can be explained by
formation of a collision complex whose decay is statistical.
\cite{Levine:1987} It has been shown that the exact quantum
scattering results for insertion reactions are in excellent
agreement with results based on a quantum statistical theory at
thermal energies. \cite{Rackham:2003}

The success of the Langevin model for total inelastic cross
sections makes it worthwhile to investigate whether the product
state distributions are well described by statistical theory. The
statistical prior distributions are given by
\begin{equation}
p_{vn} \sim (2n+1)\sqrt{E-E_{vn}},
\end{equation}
where $v$ and $n$ refer to the final states and $E$ is the total energy
in the system.

\begin{figure} [tbp]
\begin{center}
\includegraphics[width=75mm,angle=0]{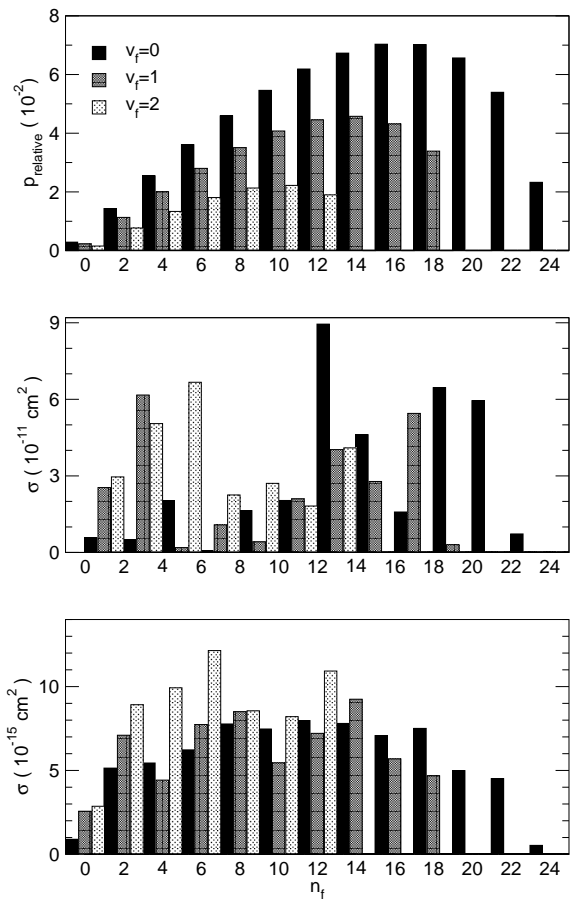}
\caption{Final rotational distributions for
$^7\mbox{Li}+\mbox{}^7\mbox{Li}_2(v_i=3,n_i=0)$. Top panel:
statistical prediction; center panel: ultracold regime (0.928 mK);
bottom panel: collision energy of 580 mK.} \label{fig18}
\end{center}
\end{figure}

Figure \ref{fig18} compares the statistical predictions with full
dynamics results in both the ultracold regime and at 580 mK for
$^7$Li + $^7$Li$_2$ collisions for initial $(v,n)=(3,0)$. The
ultracold results are quite different from the statistical
predictions and show oscillations with $n$ for each $v$ similar to
those in Fig. \ref{fig14}. However, the oscillations start to wash
out at higher energies and the rotational distributions start to
resemble the statistical ones, though they still favor low
rotational quantum numbers. The vibrational distributions are
clearly some way from statistical even at 580 mK and favor high
$v$ (smaller changes in vibrational quantum number).

\begin{figure} [tbp]
\begin{center}
\includegraphics[width=60mm,angle=-90]{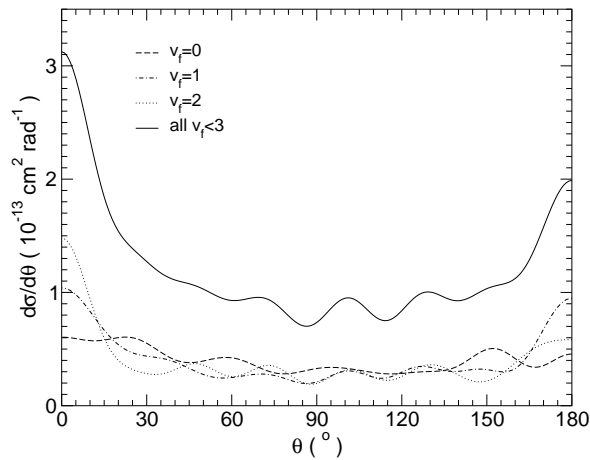}
\caption{Center-of-mass differential cross sections for
$^7\mbox{Li}+\mbox{}^7\mbox{Li}_2(v_i=3,n_i=0)$ at a collision
energy of 580 mK. Differential cross sections are integrated
through the azimuthal angle and summed over the final states in
each vibrational manifold.} \label{fig19}
\end{center}
\end{figure}

The corresponding results for differential cross sections are
shown in Fig.\ \ref{fig19}. Statistical theory predicts
forward-backward symmetry, but this has not fully developed even
at 580 mK. Similar results have been obtained in the study of cold
collisions of K + K$_2$. \cite{Quemener:2005} The symmetry
predicted by statistical theory arises due to phase cancellations
between cross terms in the sums over partial waves in Eq.\
(\ref{diffcross}), and is incomplete at energies where only a few
partial waves contribute. We expect that at higher collision
energies, where more partial waves contribute, the symmetry
observed in other deep-well systems will be recovered.

\subsection{Potential sensitivity}

Cross sections for elastic atom-atom collisions in the limit of
zero collision energy depend on a single parameter, the scattering
length. The scattering length is very sensitive to details of the
interaction potential and its properties and dependence on
potential are well established. \cite{Gribakin:1993} In
particular, the scattering length passes through a pole whenever
there is a bound state of the diatomic potential curve at exactly
zero energy, and the elastic cross section shows a corresponding
peak of height $4\pi/k^2$ (or $8\pi/k^2$ for identical bosons).

A quantitative theoretical prediction of the scattering length is
within reach today only for the lightest diatomic systems and with
the best available {\em ab initio} potentials. \cite{Gadea:2001,
Gadea:2002, Dickinson:2004} For Li$_2$, the sensitivity of
scattering length to small differences between potentials is
illustrated in Table \ref{tab1}. For heavier systems the variation
of scattering length with potential parameters is usually too fast
to allow quantitative predictions until experimental data are
available to help determine the scattering length.


The absolute accuracy of the cross sections presented in this
paper is limited by the quality of potential energy surface. The
accuracy of the electronic energies is degraded by the limitations
of the correlation treatment and the size of the basis set used.
Moreover, the effects of conical intersections and the influence
of the excited electronic state have been neglected in our
calculations. It is thus very worthwhile to investigate how
changes in the potential energy surface affect the ultracold
scattering results.

Since the potential for Li$_2$ is accurately known, we modified
the Li$_3$ potential by scaling the nonadditive part of the
potential by a multiplicative parameter $\lambda$. The depth of
the global minimum $V_{\rm min}$ varies with the scaling factor as
$\Delta V_{{\rm min}} / \Delta \lambda \approx 5150$ cm$^{-1}$
over the range studied.

\begin{figure} [tbp]
\begin{center}
\includegraphics[width=65mm,angle=-90]{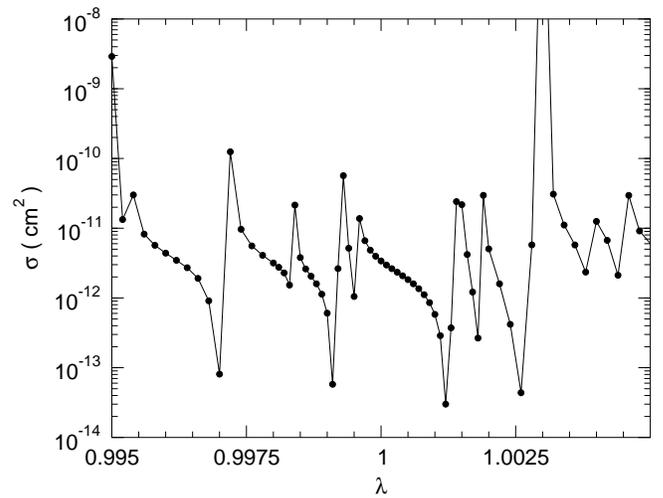}
\caption{Dependence of the elastic cross sections for
$^7\mbox{Li}+\mbox{}^7\mbox{Li}_2(v_i=0,n_i=0)$ on the scaling
factor $\lambda$ of the nonadditive part of the potential.}
\label{fig20}
\end{center}
\end{figure}

The dependence of the elastic cross sections on the scaling factor
$\lambda$ for $^7\mbox{Li} + \mbox{}^7\mbox{Li}_2(v_i=0,n_i=0)$ at
$\sim 1$ nK is shown in Figure \ref{fig20}. Whenever there is a
bound state at exactly zero energy, the scattering length passes
through a pole and the cross section shows a very high peak
($\sigma \approx 4\pi/k^2$). Near each pole is a point where
$S_{ii}=1$ so $\sigma=0$. It may be seen that the cross section
passes through about 10 cycles for a 1\% change in $\lambda$.

The mean value of the cross section can be estimated using the
formula for the mean scattering length $\overline a$ given by
Gribakin and Flambaum. \cite{Gribakin:1993} Taking the isotropic
atom-molecule dispersion coefficient $C_6=3085.54$ $E_ha_0^6$, we
obtain $\overline a = 5.83 \times 10^{-13}$ cm$^2$, which
underestimates our accurate quantum results but gives the correct
order of magnitude.

\begin{figure} [tbp]
\begin{center}
\includegraphics[width=65mm,angle=-90]{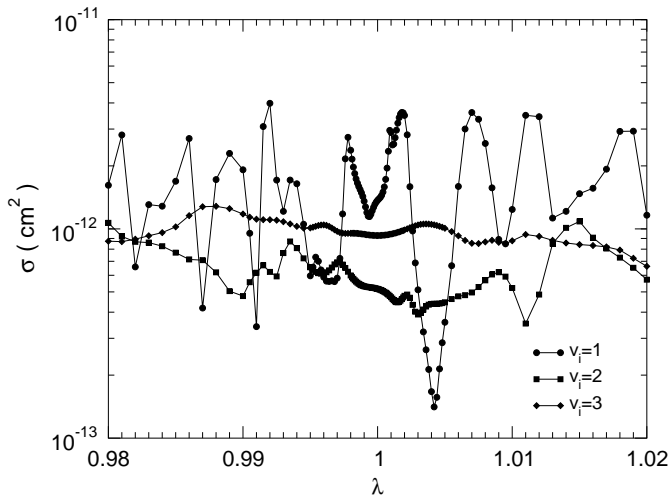}
\caption{Dependence of the elastic cross sections for
$^7\mbox{Li}+\mbox{}^7\mbox{Li}_2(v_i,n_i=0)$ on the scaling
factor $\lambda$ of the nonadditive part of the potential.}
\label{fig21}
\end{center}
\end{figure}

\begin{figure} [tbp]
\begin{center}
\includegraphics[width=65mm,angle=-90]{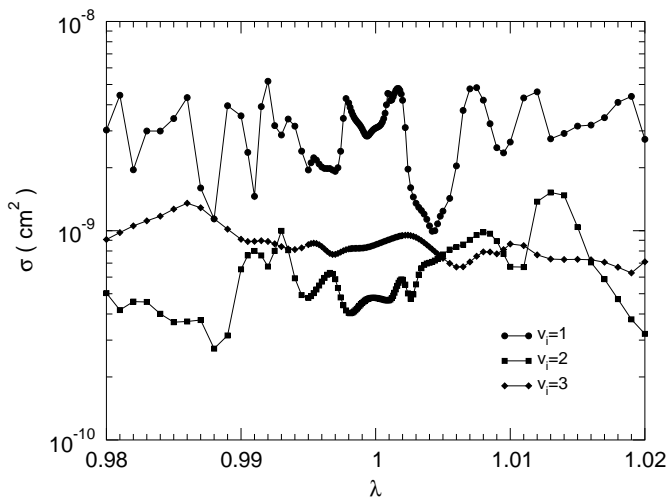}
\caption{Dependence of the total inelastic cross sections for
$^7\mbox{Li}+\mbox{}^7\mbox{Li}_2(v_i,n_i=0)$ on the scaling
factor $\lambda$ of the nonadditive part of the potential.}
\label{fig22}
\end{center}
\end{figure}

For collisions of molecules in vibrationally excited states
$(v_i,n_i=0)$ with $v_i=1$ to 3, the dependence of elastic and
inelastic cross sections at $\sim 1$ nK on $\lambda$ is shown in
Figures \ref{fig21} and \ref{fig22}. The dependence is again
oscillatory, but the frequency and amplitude of the oscillations
are smaller for higher initial vibrational quantum numbers $v_i$.
The elastic (inelastic) cross sections vary across the range by
186\% (136\%) for $v_i=1$ and 64\% (73\%) for $v_i=3$. The sharp
peaks in cross sections associated with poles in scattering
lengths are entirely washed out for higher values of $v_i$.


The lack of strong resonant peaks as a function of well depth for
Li + Li$_2$ is an example of a general effect recently discussed
by Hutson. \cite{Hutson:res:2007} Similar effects have recently
been observed when tuning through zero-energy Feshbach resonances
in He + NH using magnetic fields. \cite{Gonzalez-Martinez:2007} In
the presence of inelastic scattering, the complex scattering
length $a=\alpha-{\rm i}\beta$ exhibits an oscillation rather than
a pole when a bound-state crosses threshold. The amplitude of the
oscillation is characterized by a {\it resonant scattering length}
$a_{\rm res}$. If the bound state is coupled with comparable
strengths to the elastic and inelastic channels, the oscillation
may be of relatively low amplitude and the peaks in cross sections
are suppressed. This is the case in Li + Li$_2$, where there are
strong couplings between all the channels involved. For $v_i=1$
there is enough resonant coupling to the elastic channel that
peaks are still observed, but even these have an amplitude of only
about a factor of 10. For higher $v_i$, there are larger numbers
of inelastic channels, each of which competes with the elastic
channel and the resonant structures are progressively washed out.

The suppression of cross-section peaks observed here for Li +
Li$_2$ contrasts with the situation commonly found for atom-atom
scattering \cite{Koehler:RMP:2006} and for atom-molecule reactive
scattering in systems such as F + H$_2$. \cite{Bodo:FH2:2004} For
atom-atom collisions involving atoms in spin-stretched states, the
resonant state is coupled to inelastic channels only by very week
magnetic dipole-dipole interactions, but is coupled to the elastic
channel by much stronger spin-exchange terms. This gives a large
value of $a_{\rm res}$, corresponding to pole-like behavior of the
scattering length and producing very large peaks in cross
sections. Similarly, in F + H$_2$ the resonances studied
\cite{Bodo:FH2:2004} are due to quasibound states in the entrance
channel, while the ``inelastic'' channels are actually reactive
channels that are separated from the entrance well by a high
barrier. This again gives very large peaks in cross sections. It
is evident that the factor that suppresses the resonant peaks in
Li + Li$_2$ is strong coupling to the inelastic channels.

A sensitivity study similar to the one we report here was
performed on the $\mbox{Na}+\mbox{Na}_2$ system by Qu\'em\'ener
{\em et al.} \cite{Quemener:2004} They also observed oscillations
that decreased in amplitude with initial $v$.

\subsection{Isotope effects}

A strong suppression of inelastic rates has been observed
experimentally in fermionic systems tuned to a large and positive
atom-atom scattering length. \cite{Strecker:2003, Cubizolles:2003,
Jochim:Li2pure:2003, Regal:lifetime:2004} The molecular cloud
consisting of weakly bound dimers of fermions exhibits remarkable
stability against collisional decay. By contrast, the vibrational
relaxation for weakly bound bosonic dimers is fast.
\cite{Stenger:1999, Donley:2002, Mukaiyama:2004, Herbig:2003,
Durr:mol87Rb:2004} The suppression of inelastic scattering for
fermionic systems has been interpreted in terms of the
requirements of Fermi statistics. \cite{Petrov:2004,
Petrov:suppress:2005}.

We have repeated all the above scattering calculations involving
molecules in low-lying rovibrational states for the fermionic
system $^6$Li + $^6$Li$_2$. A brief account of the results was
given in ref.~\onlinecite{Cvitas:bosefermi:2005}. For $^6$Li +
$^6$Li$_2$, the dominant contribution to cross sections in the
ultracold regime comes from total angular momentum $J=1^{-}$,
which is the partial wave that contains a contribution from
$l_i=0$. Cross sections at a collision energy of $\approx 1$ nK
are shown in Table \ref{crossrate_fer}. The individual calculated
values are significantly different from the bosonic case, but the
effect is {\it not} due to suppression by Fermi statistics.
Instead, the dominant effects is that the change in nuclear mass
alters the scattering by a mechanism similar to a change in the
potential surface, as discussed in the previous subsection. The
cross sections calculated for individual initial states are thus
once again scattered essentially randomly about a mean value.
However, when we compare the results for a variety of initial
rovibrational states, we see that there are no {\it systematic}
differences between the overall rates for bosonic and fermionic
systems for the low vibrational states that we have studied.


Unusually low inelastic rates (about an order of magnitude lower,
with small $\mbox{Im}(a)$) were obtained for systems that have a
small number of energetically accessible inelastic channels, such
as $^6\mbox{Li}+ \mbox{}^6\mbox{Li}_2(v_i=0,n_i=3)$ and
$^7\mbox{Li}+ \mbox{}^7\mbox{Li}_2(v_i=0,n_i=2)$.

\begin{table}[btp]
\begin{center}
\begin{tabular}{cccc} \hline
$v_i$, $n_i$ \quad&\quad $\sigma_{{\rm elas}}$ (cm$^{2}$)\quad
&\quad $\sigma_{{\rm inel}}$ (cm$^{2}$) \quad &
$\sigma_{{\rm inel}}/\sigma_{{\rm elas}}$ \\
\hline
0, 1  & $2.20 \times 10^{-12}$ & $-$ & $-$ \\
0, 3  & $1.11 \times 10^{-12}$ & $2.62 \times 10^{-10}$ & 236 \\
0, 5  & $9.27 \times 10^{-13}$ & $2.17 \times 10^{-9}$ & 2340 \\
0, 7  & $1.32 \times 10^{-13}$ & $8.90 \times 10^{-10}$ & 6740 \\
0, 9  & $1.07 \times 10^{-12}$ & $1.81 \times 10^{-9}$  & 1690 \\
0, 11 & $1.77 \times 10^{-12}$ & $2.37 \times 10^{-9}$  & 1340 \\
\hline
1, 1  & $6.09 \times 10^{-13}$ & $1.40 \times 10^{-9}$ & 2300 \\
1, 3  & $1.21 \times 10^{-12}$ & $1.21 \times 10^{-9}$ & 1000 \\
1, 5  & $1.93 \times 10^{-12}$ & $2.59 \times 10^{-9}$ & 1340 \\
1, 7  & $1.43 \times 10^{-12}$ & $2.02 \times 10^{-9}$ & 1410 \\
1, 9  & $1.74 \times 10^{-12}$ & $1.96 \times 10^{-9}$ & 1130 \\
1, 11 & $1.27 \times 10^{-12}$ & $2.31 \times 10^{-9}$ & 1820 \\
\hline
2, 1  & $1.67 \times 10^{-12}$ & $2.26 \times 10^{-9}$ & 1350 \\
2, 3  & $1.14 \times 10^{-12}$ & $2.23 \times 10^{-9}$  & 1960 \\
2, 5  & $2.04 \times 10^{-12}$ & $2.80 \times 10^{-9}$  & 1370 \\
2, 7  & $1.00 \times 10^{-12}$ & $2.52 \times 10^{-9}$  & 2520 \\
2, 9  & $1.46 \times 10^{-12}$ & $3.20 \times 10^{-9}$  & 2190 \\
2, 11 & $7.72 \times 10^{-13}$ & $1.77 \times 10^{-9}$  & 2290 \\
\hline
3, 1  & $1.46 \times 10^{-12}$ & $2.76 \times 10^{-9}$ & 1890 \\
3, 3  & $2.62 \times 10^{-12}$ & $2.57 \times 10^{-9}$  & 981 \\
3, 5  & $1.87 \times 10^{-12}$ & $2.71 \times 10^{-9}$  & 1450 \\
3, 7  & $1.31 \times 10^{-12}$ & $2.77 \times 10^{-9}$  & 2110 \\
3, 9  & $1.17 \times 10^{-12}$ & $2.10 \times 10^{-9}$  & 1790 \\
3, 11 & $6.92 \times 10^{-13}$ & $2.00 \times 10^{-9}$  & 2890 \\
\cline{1-4}
\end{tabular}
\vspace{0.5cm} \caption{Elastic and total inelastic cross sections
and rate coefficients for $^6\mbox{Li} +
\mbox{}^6\mbox{Li}_2(v_i,n_i)$ at a collision energy of 0.928 nK
for different initial states of the molecule.}
\label{crossrate_fer}
\end{center}
\end{table}

The energy dependence of the cross sections in fermionic systems
is qualitatively similar to that described above for bosonic
systems. The Wigner regime is reached at energies around 10$^{-5}$
K. For non-zero initial end-over-end angular momenta, the simple
relationship of Eq.\ \ref{wig_laws} holds at energies below the
corresponding centrifugal barriers. At higher energies, the energy
dependence of inelastic cross sections follows the Langevin model,
Eq.\ \ref{cross_langevin}.

\begin{figure} [tbp]
\begin{center}
\includegraphics[width=60mm,angle=-90]{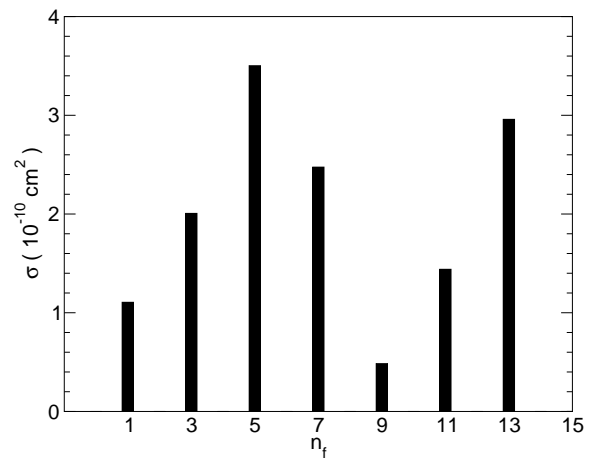}
\caption{Final rotational distributions for
$^6\mbox{Li}+\mbox{}^6\mbox{Li}_2(v_i=1,n_i=1)$ at a collision
energy in the Wigner regime.} \label{fig23}
\end{center}
\end{figure}

\begin{figure} [tbp]
\begin{center}
\includegraphics[width=60mm,angle=-90]{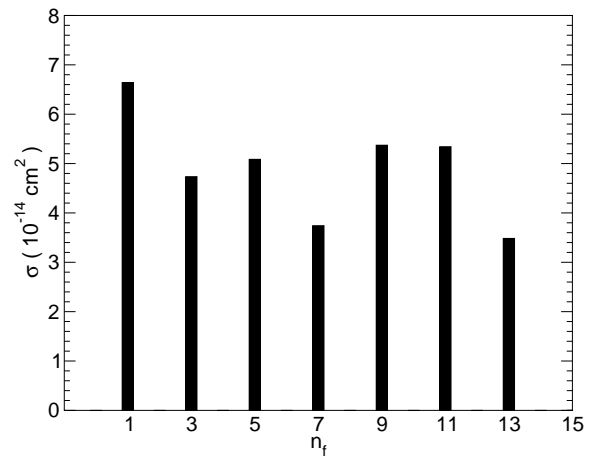}
\caption{Final rotational distributions for
$^6\mbox{Li}+\mbox{}^6\mbox{Li}_2(v_i=1,n_i=1)$ at a collision
energy of 116 mK.} \label{fig24}
\end{center}
\end{figure}

\begin{figure} [tbp]
\begin{center}
\includegraphics[width=60mm,angle=-90]{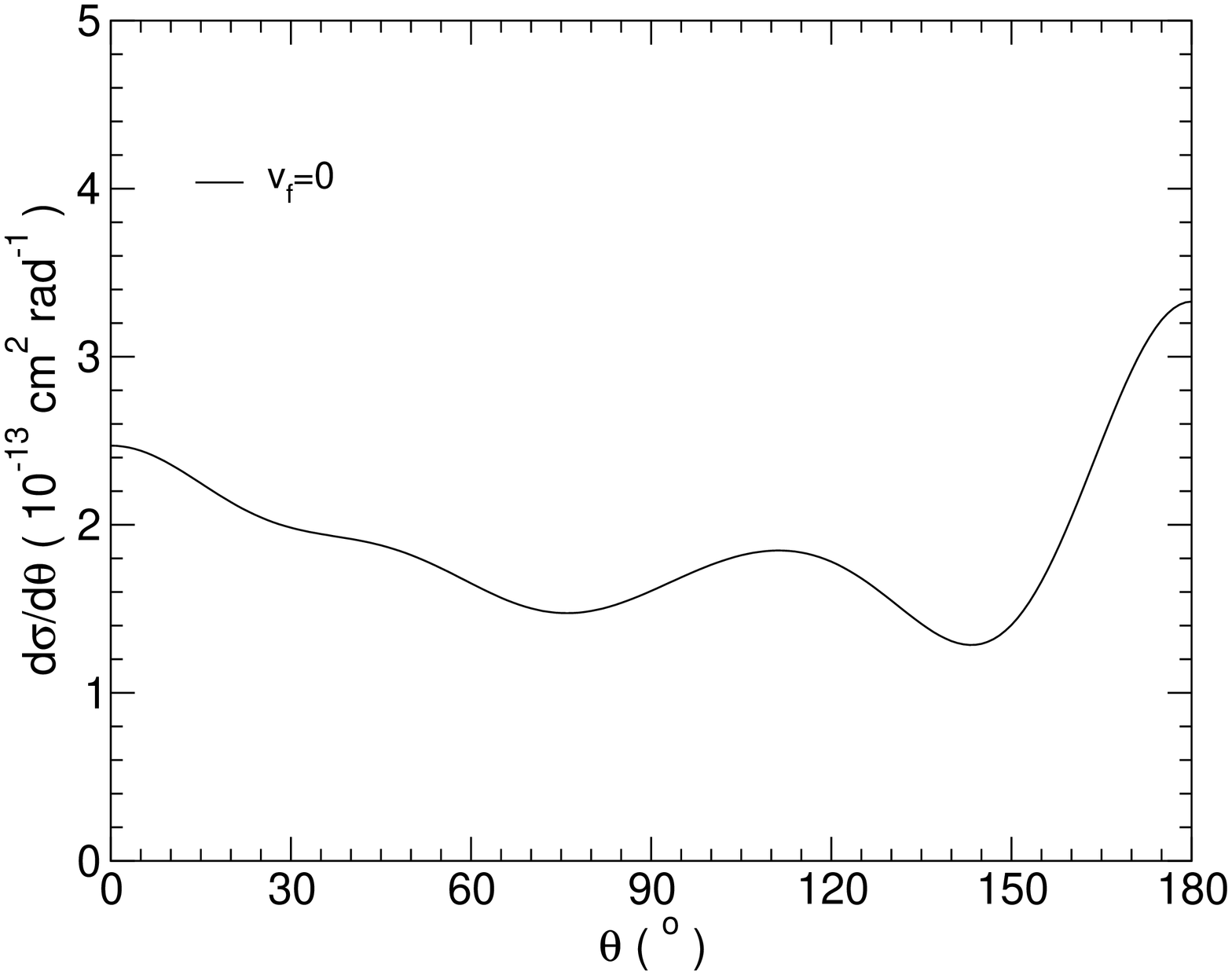}
\caption{Center-of-mass differential cross sections for
$^6\mbox{Li}+\mbox{}^6\mbox{Li}_2(v_i=1,n_i=1)$ at a collision
energy of 116 mK. Differential cross sections are integrated
through the azimuthal angle and summed over the final states in
each vibrational manifold and overall.} \label{fig25}
\end{center}
\end{figure}

For spin-stretched $^6$Li + $^6$Li$_2$ collisions, only odd
rotational levels of Li$_2$ are allowed. Figure \ref{fig23} shows
the product rotational distribution for $v_i=1$, $n_i=1$ in the
Wigner regime, and it may be seen that the distribution is again
oscillatory. At higher collision energies the oscillations are
washed out in the sum over partial waves, as shown in Figure
\ref{fig24}. However, as for the bosonic system, the rotational
distribution is significantly non-statistical. The
forward-backward symmetry of the differential cross sections
predicted by statistical theory is also not present for $^6$Li +
$^6$Li$_2$ at 116 mK, as seen in Figure \ref{fig25}.

\subsection{Reactions in isotopic mixtures}

\begin{figure}[tbp]
\begin{center}
\includegraphics[width=75mm,angle=0]{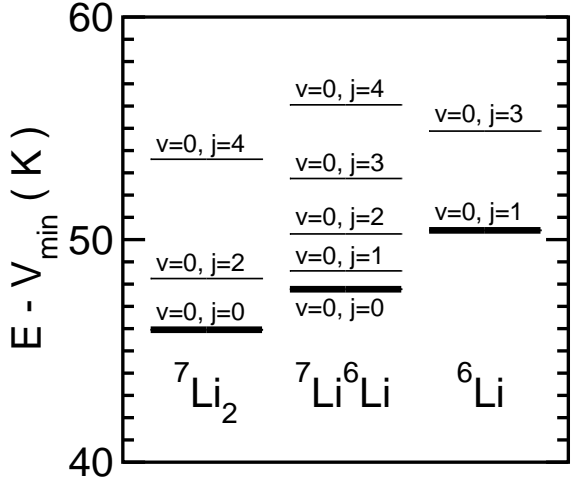}
\caption{The low-lying energy rotational levels of $^7$Li$_2$,
$^6$Li$^7$Li and $^6$Li$_2$ for $v=0$, relative to the Li$_2$
potential minimum. Only levels with even $n$ are shown for
$^7$Li$_2$ and only levels with odd $n$ are shown for $^6$Li$_2$.}
\label{fig26}
\end{center}
\end{figure}

A greater variety of atom--diatom collisions may be observed in
mixtures. Isotopic mixtures of $^6$Li and $^7$Li are of
considerable interest, \cite{Truscott:2001, Schreck:2001,
Khaykovich:2003} and one can envisage creating either homonuclear
or heteronuclear dimers in such a mixture. The atom-diatom
collision systems of interest are
\begin{equation}
^7\mbox{Li}+\mbox{}^6\mbox{Li}\mbox{}^7\mbox{Li},
\label{li_bfb}
\end{equation}
\begin{equation}
^7\mbox{Li}+\mbox{}^6\mbox{Li}_2. \label{li_bff}
\end{equation}
\begin{equation}
^6\mbox{Li}+\mbox{}^7\mbox{Li}_2,
\label{li_fbb}
\end{equation}
\begin{equation}
^6\mbox{Li}+\mbox{}^6\mbox{Li}\mbox{}^7\mbox{Li},
\label{li_ffb}
\end{equation}
We have studied all four of the systems (\ref{li_bfb}) to
(\ref{li_ffb}) and gave a preliminary report in ref.\
\onlinecite{Cvitas:hetero:2005}. The novel feature in these
systems is the possibility of a chemical reaction in which the
reactants and products are distinguishable. The low-lying energy
levels of $^7$Li$_2$, $^6$Li$^7$Li and $^6$Li$_2$ are shown
relative to the Li$_2$ potential minimum in Fig.\ \ref{fig26}.
Because of the differences in zero-point energy, the reactions are
exoergic even for ground-state molecules for the two systems
involving atomic $^7$Li, (\ref{li_bfb}) and (\ref{li_bff}).

For $^6$Li$^7$Li, both even and odd rotational levels are allowed
even for spin-stretched states. The higher density of states makes
these systems harder to treat computationally. Our calculations on
these systems were therefore restricted to the $J=0$ and 1 partial
waves and are thus converged only at ultracold collision energies.

We will consider the systems that are reactive for ground-state
molecules first. The cross sections, rate coefficients and
scattering lengths for collisions of $^7$Li with $^6$Li$^7$Li in
several different initial molecular states are summarized in Table
\ref{crossrate_bfb}. For molecules in excited vibrational or
rotational states, the cross sections are comparable to those
found for the homonuclear systems. However, for
$^6$Li$^7$Li($v=0,n=0$) the cross section for reaction is a factor
of about 50 smaller than for other states. This corresponds to a
reactive rate coefficient of $4.05 \times 10^{-12}$ cm$^3$
s$^{-1}$ and Im$(a)=0.233$ \AA. In this case there is only a
single exoergic channel, with a kinetic energy release of 1.822 K.
The small cross section is probably due to the small volume of
available phase space for reaction.

\begin{table*}[tbp]
\begin{center}
\begin{tabular}{ccccccc} \hline
$v_i$, $n_i$ \quad&\quad $\sigma_{{\rm elas}}$ (cm$^{2}$)\quad
&\quad $\sigma_{{\rm inel}}$ (cm$^{2}$) \quad & \quad
$\sigma_{{\rm reac}}$ (cm$^{2}$)\quad &\quad $k_{{\rm loss}}$
(cm$^{3}$ s$^{-1}$)
\quad & \quad Re($a$) (nm)\quad & \quad $-$Im($a$) (nm)\quad \\
\hline
0, 0  & $1.77 \times 10^{-13}$ & $-$ & $2.20 \times 10^{-11}$ &
$4.05 \times 10^{-12}$ & 1.19 & 0.0233 \\
1, 0  & $1.32 \times 10^{-12}$ & $8.79 \times 10^{-10}$ & $2.56 \times 10^{-10}$ &
$2.09 \times 10^{-10}$ & 3.01 & 1.19 \\
2, 0  & $1.32 \times 10^{-12}$ & $1.37 \times 10^{-9}$ & $1.02 \times 10^{-9}$ &
$4.40 \times 10^{-10}$ & 2.04 & 2.52 \\
3, 0  & $1.09 \times 10^{-12}$ & $1.30 \times 10^{-9}$ & $8.74 \times 10^{-10}$ &
$4.00 \times 10^{-10}$ & 1.86 & 2.29 \\
\hline
0, 1  & $4.81 \times 10^{-12}$ & $4.30 \times 10^{-10}$ & $2.32 \times 10^{-9}$ &
$5.06 \times 10^{-10}$ & 5.46 & 2.90 \\
1, 1  & $1.52 \times 10^{-12}$ & $1.85 \times 10^{-9}$ & $9.47 \times 10^{-10}$ &
$5.14 \times 10^{-10}$ & 1.85 & 2.94 \\
2, 1  & $9.09 \times 10^{-13}$ & $1.22 \times 10^{-9}$ & $4.45 \times 10^{-10}$ &
$2.93 \times 10^{-10}$ & 2.10 & 1.68 \\
3, 1  & $9.51 \times 10^{-13}$ & $1.67 \times 10^{-9}$ & $6.28 \times 10^{-10}$ &
$4.23 \times 10^{-10}$ & 1.71 & 2.15 \\
\cline{1-7}
\end{tabular}
\vspace{0.5cm} \caption{Cross sections and related parameters for
$^7$Li + $^6$Li$^7$Li $(v_i,n_i)$ at a collision energy of 0.928
nK for different initial states of the molecule.}
\label{crossrate_bfb}
\end{center}
\end{table*}

\begin{figure}[tbp]
\begin{center}
\includegraphics[width=75mm,height=115mm,angle=0]{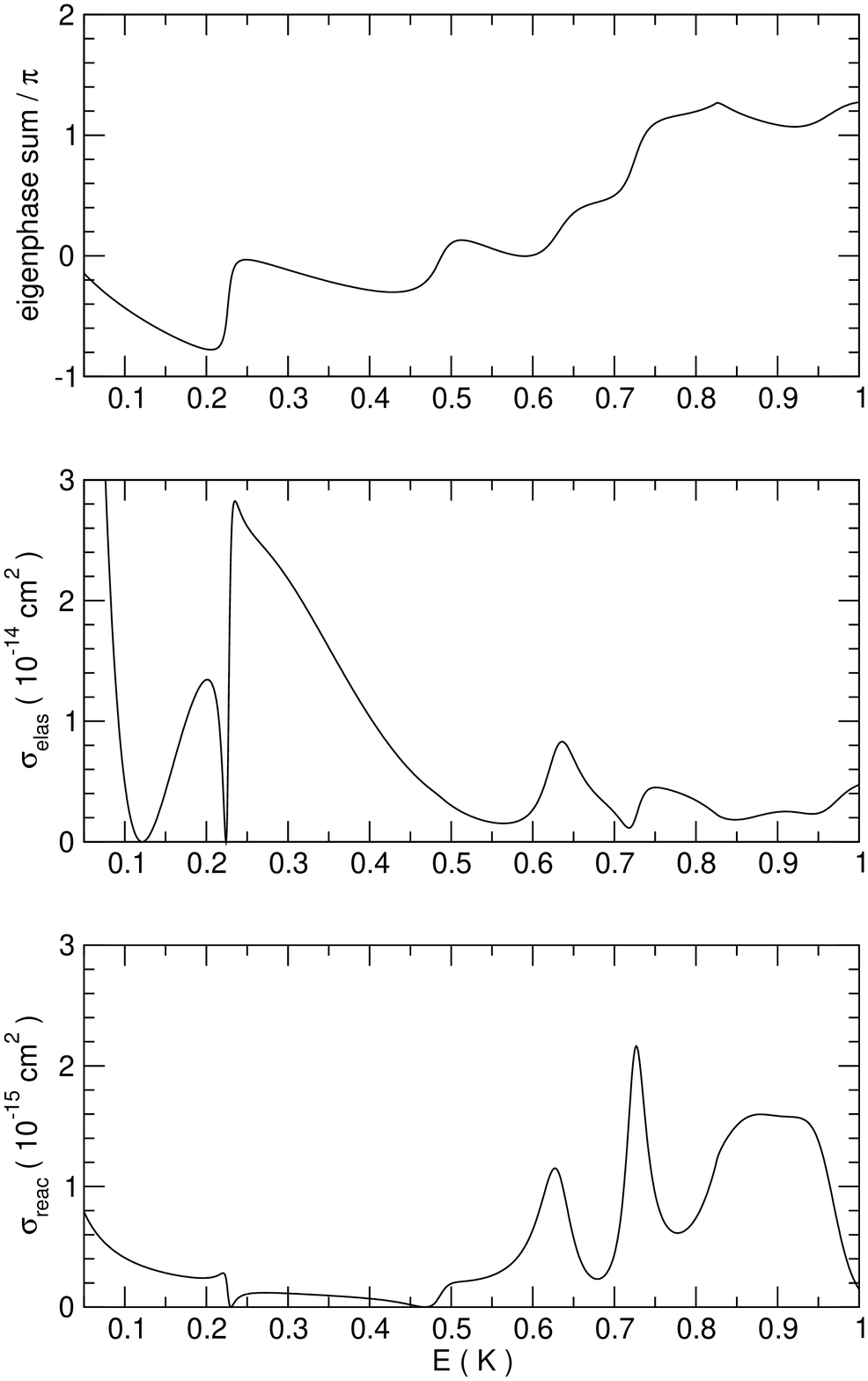}
\caption{Energy dependence of eigenphase sum (top panel), elastic
cross section (center panel) and inelastic/reactive cross section
(bottom panel) for collisions of $^6$Li$^7$Li ($v=0,n=0$) with
$^7$Li.} \label{fig27}
\end{center}
\end{figure}

The increased density of rotational states for $^6$Li$^7$Li
produces an increased density of Feshbach resonances. The
energy-dependence of the eigenphase sum for $^7$Li + $^6$Li$^7$Li
is shown in the top panel of Figure \ref{fig27}. There is a
Feshbach resonance clearly visible at 225 mK and another at about
470 mK, though the latter overlaps the threshold for opening the
$n=2$ reactive channel at 477.2 mK. There are further overlapping
resonances near 630 mK and 725 mK. The corresponding structures in
the elastic and inelastic s-wave cross sections are shown in the
lower two panels of Figure \ref{fig27}. The elastic cross section
shows a minimum near 120 mK, where the phase shift obtained from
the elastic S-matrix element passes through a multiple of $\pi$.
It dips to a very small value near the Feshbach resonance at 225
mK, but {\it not} near the resonance near 470 mK. The reactive
cross section, by contrast, dips close to zero at {\it both}
Feshbach resonances. This suggests the interesting possibility of
reducing the inelastic/elastic cross section ratio by tuning close
to a Feshbach resonance.

Reactive collisions of $^7$Li with $^6$Li$_2$ are more strongly
exoergic because for spin-stretched states the $^6$Li$_2$ molecule is
restricted to odd rotational levels while the $^6$Li$^7$Li
molecule is not. Because of this collisions of $^7$Li with
$^6$Li$_2$ in its $n=1$ ground state can produce $^6$Li$^7$Li in
$n=0$, 1 or 2. The calculated cross sections, rate coefficients
and scattering lengths are summarized in Table
\ref{crossrate_bff}. The reactive cross sections for $^6$Li$_2$
($n=1$) are again reduced compared to those for other states, but
not as dramatically as for $^7$Li + $^6$Li$^7$Li ($n=0$).

\begin{table*}[btp]
\begin{center}
\begin{tabular}{ccccccc} \hline
$v_i$, $n_i$ \quad&\quad $\sigma_{{\rm elas}}$ (cm$^{2}$)\quad
&\quad $\sigma_{{\rm inel}}$ (cm$^{2}$) \quad & \quad
$\sigma_{{\rm reac}}$ (cm$^{2}$)\quad &\quad $k_{{\rm loss}}$
(cm$^{3}$ s$^{-1}$)
\quad & \quad Re($a$) (nm)\quad & \quad $-$Im($a$) (nm)\quad \\
\hline
0, 1  & $1.17 \times 10^{-12}$ &                        & $2.34 \times 10^{-10}$ &
$4.37 \times 10^{-11}$ & 3.04 & 0.244 \\
1, 1  & $1.62 \times 10^{-12}$ & $1.00 \times 10^{-9}$ & $1.76 \times 10^{-9}$ &
$5.15 \times 10^{-10}$ & 2.17 & 2.86 \\
2, 1  & $6.61 \times 10^{-13}$ & $2.66 \times 10^{-10}$ & $1.12 \times 10^{-9}$ &
$2.58 \times 10^{-10}$ & 1.41 & 1.81 \\
3, 1  & $8.05 \times 10^{-13}$ & $5.29 \times 10^{-10}$ & $1.07 \times 10^{-9}$ &
$2.98 \times 10^{-10}$ & 1.91 & 1.66 \\
\cline{1-7}
\end{tabular}
\vspace{0.5cm} \caption{Cross sections and related parameters for
$^7$Li + $^6$Li$_2$ $(v_i,n_i)$ at a collision energy of 0.928 nK
for different initial states of the molecule.}
\label{crossrate_bff}
\end{center}
\end{table*}

Ultracold collisions of $^6$Li$^7$Li and $^7$Li$_2$ with atomic
$^6$Li are nonreactive unless the molecule is initially in a
rovibrationally excited state. This is because for these systems
the zero-point energy is greater in the products than in the
reactants. The calculated cross sections, rate coefficients and
scattering lengths for these systems are summarized in Tables
\ref{crossrate_ffb} and \ref{crossrate_fbb}. The patterns of cross
sections for rovibrationally excited initial states are fairly
similar to those for the homonuclear systems, with no particularly
small inelastic or reactive rates. The novel feature of these
systems really lies in the fact that the reactions are forbidden
for ground-state molecules, so that sympathetic cooling of
molecules by contact with $^6$Li atoms might be envisaged. For
example, if $^6$Li$_2$ molecules are produced in the presence of
both $^6$Li and $^7$Li, they can react to form (relatively hot)
$^6$Li$^7$Li. If the $^7$Li atoms are then removed, the molecules
might be cooled by elastic collisions with $^6$Li.

\begin{table*}[btp]
\begin{center}
\begin{tabular}{ccccccc} \hline
$v_i$, $n_i$ \quad&\quad $\sigma_{{\rm elas}}$ (cm$^{2}$)\quad
&\quad $\sigma_{{\rm inel}}$ (cm$^{2}$) \quad & \quad
$\sigma_{{\rm reac}}$ (cm$^{2}$)\quad &\quad $k_{{\rm loss}}$
(cm$^{3}$ s$^{-1}$)
\quad & \quad Re($a$) (nm)\quad & \quad $-$Im($a$) (nm)\quad \\
\hline
0, 0  & $4.71 \times 10^{-12}$ &  &  &  & 6.12 &  \\
1, 0  & $6.43 \times 10^{-13}$ & $1.15 \times 10^{-9}$ & $2.12 \times 10^{-10}$ &
$2.64 \times 10^{-10}$ & 1.81 & 1.36 \\
2, 0  & $1.12 \times 10^{-12}$ & $1.39 \times 10^{-9}$ & $4.37 \times 10^{-10}$ &
$3.53 \times 10^{-10}$ & 2.36 & 1.82 \\
3, 0  & $1.54 \times 10^{-12}$ & $1.34 \times 10^{-9}$ & $9.49 \times 10^{-10}$ &
$4.43 \times 10^{-10}$ & 2.66 & 2.29 \\
\hline
0, 1  & $2.74 \times 10^{-12}$ & $5.84 \times 10^{-10}$ &  &
$1.13 \times 10^{-10}$ & 4.64 & 0.586 \\
1, 1  & $7.56 \times 10^{-13}$ & $2.04 \times 10^{-9}$ & $2.72 \times 10^{-10}$ &
$4.48 \times 10^{-10}$ & 0.828 & 2.31 \\
2, 1  & $1.19 \times 10^{-12}$ & $2.07 \times 10^{-9}$ & $7.21 \times 10^{-10}$ &
$5.40 \times 10^{-10}$ & 1.32 & 2.79 \\
3, 1  & $1.27 \times 10^{-12}$ & $1.55 \times 10^{-9}$ & $7.76 \times 10^{-10}$ &
$4.51 \times 10^{-10}$ & 2.16 & 2.33 \\
\cline{1-7}
\end{tabular}
\vspace{0.5cm} \caption{Cross sections and related parameters for
$^6$Li + $^6$Li$^7$Li $(v_i,n_i)$ at a collision energy of 0.928
nK for different initial states of the molecule.}
\label{crossrate_ffb}
\end{center}
\end{table*}

\begin{table*}[btp]
\begin{center}
\begin{tabular}{ccccccc} \hline
$v_i$, $n_i$ \quad&\quad $\sigma_{{\rm elas}}$ (cm$^{2}$)\quad
&\quad $\sigma_{{\rm inel}}$ (cm$^{2}$) \quad & \quad
$\sigma_{{\rm reac}}$ (cm$^{2}$)\quad &\quad $k_{{\rm loss}}$
(cm$^{3}$ s$^{-1}$)
\quad & \quad Re($a$) (nm)\quad & \quad $-$Im($a$) (nm)\quad \\
\hline
0, 0  & $1.29 \times 10^{-12}$ &  &  &  & 3.20 &  \\
1, 0  & $1.33 \times 10^{-12}$ & $5.70 \times 10^{-10}$ &  $8.85 \times 10^{-10}$ &
$2.79 \times 10^{-10}$ & 2.90 & 1.47 \\
2, 0  & $1.14 \times 10^{-12}$ & $1.39 \times 10^{-9}$ &  $1.37 \times 10^{-9}$ &
$5.28 \times 10^{-10}$ & 1.16 & 2.79 \\
3, 0  & $1.51 \times 10^{-12}$ & $1.00 \times 10^{-9}$ &  $1.42 \times 10^{-9}$ &
$4.63 \times 10^{-10}$ & 2.45 & 2.45 \\
\cline{1-7}
\end{tabular}
\vspace{0.5cm} \caption{Cross sections and related parameters for
$^6$Li + $^7$Li$_2$ $(v_i,n_i)$ at a collision energy of 0.928 nK
for different initial states of the molecule.}
\label{crossrate_fbb}
\end{center}
\end{table*}

\section{Conclusions}

We have developed a new potential energy surface for quartet Li +
Li$_2$ from high-level electronic structure calculations and used
it to carry out quantum dynamics calculations on elastic,
inelastic and reactive collisions on Li + Li$_2$ under cold and
ultracold collisions. The potential energy surface was calculated
using RCCSD(T) calculations with all electrons correlated. The
surface was calculated on a grid in pure bond-length coordinates.
It was interpolated at short range using the interpolant-moving
least-squares method and merged with a long-range form that is
correct at both the atom-diatom and the atom-atom-atom
dissociation limits.

The potential energy surface has a deep triangular minimum that
lies far below the atom-diatom energy and allows barrierless atom
exchange reactions. The well depth is a factor of 4 deeper than
would be expected on the basis of a pairwise-additive sum of
triplet Li$_2$ potentials. The surface exhibits a seam of conical
intersections at linear geometries, at energies close to that at
three separated Li atoms.

Quantum dynamics calculations were carried out in a
time-independent reactive scattering formalism based on an
expansion in pseudohyperspherical harmonics. This approach makes
it straightforward to include the required symmetry with respect
to exchange of identical bosons ($^7$Li atoms) or fermions ($^6$Li
atoms). For the homonuclear systems $^7$Li + $^7$Li$_2$ and $^6$Li
+ $^6$Li$_2$, the products of reactive and non-reactive collisions
are indistinguishable. Nevertheless, all collisions involving
vibrationally excited Li$_2$ result in very fast vibrational
relaxation (quenching). For the low-lying vibrationally excited
states studied here, there is {\it no} systematic suppression of
quenching rates for collisions involving fermion dimers
($^6$Li$_2$). This contrasts with the situation found both
experimentally and theoretically for fermion dimers produced by
Feshbach resonance tuning in their highest vibrational state.

For collisions involving mixtures of Li isotopes, reactive and
nonreactive collision outcomes can be distinguished. In
particular, for collisions of $^7$Li atoms with either $^6$Li$_2$
or $^6$Li$^7$Li, exoergic reactive collisions are possible even
for vanishing kinetic energy because of the difference in
zero-point energy between reactants and products. These reactive
processes are generally very fast, except for the case of $^7$Li +
$^6$Li$^7$Li ($v=0,n=0$) at collision energies below 477 mK, for
which there is only one reactive channel.

The Li + Li$_2$ collision systems have a rich structure of
scattering resonances at low energies. These are principally
rotational Feshbach resonances due to quasibound states that
correlate with rotationally excited diatomic molecules. Tuning
such Feshbach resonances with applied magnetic fields offers
possibilities for {\it controlling} ultracold molecular
collisions, and this will be a topic for future work.

We have investigated the sensitivity of elastic and inelastic
cross sections to variations in the interaction potential. Such
variations tune Feshbach resonances across thresholds, and our
expectation was that the cross sections would show very large
peaks when this occurred. This is indeed the case for collisions
of ground-state molecules where only elastic scattering is
possible. For such collisions the elastic cross sections increase
by many orders of magnitude close to each resonance, and the
density of states is such that a 1\% change in the potential depth
sweeps about 10 resonances across threshold. However, for
collisions where inelastic scattering is also possible the
resonant behavior is strongly suppressed. The amplitude of the
oscillations in cross sections decreases dramatically with
increasing initial vibrational quantum number, and for $v=3$ the
cross section changes by no more than a factor of 2 as resonances
cross threshold. This shows that, in some cases, cross sections
are much less sensitive to small variations in the potential than
was previously expected.

\section*{Acknowledgments}
MTC is grateful for sponsorship from the University of Durham and
Universities UK.

\bibliography{../../all}

\end{document}